\documentclass[12pt,journal,onecolumn,romanappendice]{IEEEtran}

\usepackage{graphicx,caption,subcaption,colordvi,psfrag}
\usepackage{epstopdf}
\usepackage{calc,pstricks, pgf, xcolor}
\usepackage{cite}
\usepackage{amssymb}
\usepackage{epsfig}
\usepackage{amsmath}
\usepackage{amsthm}
\usepackage{setspace}
\newtheorem{thm}{Theorem}%[chapter]
\newtheorem{lem}{Lemma}%[chapter]
\newtheorem{defn}{Definition}%[chapter]
\newtheorem{rem}{Remark}
\newtheorem{example}{Example}

\newcommand{\abs}[1]{\left\vert#1\right\vert}

\makeatletter

\newcommand{\Rmnum}[1]{\expandafter\@slowromancap\romannumeral #1@}
\makeatother

\makeatletter
\renewcommand*\env@matrix[1][c]{\hskip -\arraycolsep
  \let\@ifnextchar\new@ifnextchar
  \array{*\c@MaxMatrixCols #1}}
\makeatother

\usepackage{lscape}

\begin{document}

%\title{The Under Addressed MIMO Optical Channel: Capacity, Outage and DMT}
\title{The Jacobi MIMO Channel}

\author{\IEEEauthorblockN{Ronen Dar\IEEEauthorrefmark{1}, Meir Feder\IEEEauthorrefmark{2} and Mark Shtaif\IEEEauthorrefmark{3}}\\
\IEEEauthorblockA{School of Electrical Engineering\\
Tel Aviv University\\
Tel Aviv 69978, Israel \\
Email: \IEEEauthorrefmark{1}ronendar@post.tau.ac.il, \IEEEauthorrefmark{2}meir@eng.tau.ac.il, \IEEEauthorrefmark{3}shtaif@tauex.tau.ac.il}
}

\maketitle

\begin{abstract}
This paper presents a new fading model for MIMO channels, the Jacobi fading model. It asserts that $\mathbf H$, the transfer matrix which couples the $m_t$ inputs into $m_r$ outputs, is a sub-matrix of an $m\times m$ random (Haar-distributed) unitary matrix.
The (squared) singular values of $\mathbf H$ follow the law of the classical Jacobi ensemble of random matrices; hence the name of the channel.
One motivation to define such a channel comes from multimode/multicore optical fiber communication. It turns out that this model can be qualitatively different than the Rayleigh model, leading to interesting practical and theoretical results. This work first evaluates the ergodic capacity of the channel. Then, it considers the non-ergodic case, where it analyzes the outage probability and the diversity-multiplexing tradeoff. In the case where $k=m_t+m_r-m > 0$ it is shown that at least $k$ degrees of freedom are guaranteed not to fade for any channel realization, enabling a zero outage probability or infinite diversity order at the corresponding rates. A simple scheme utilizing (a possibly outdated) channel state feedback is provided, attaining the no-outage guarantee. Finally, noting that as $m$ increases, the Jacobi model approaches the Rayleigh model, the paper discusses the applicability of the model in other communication scenaria.
\end{abstract}

\section{Introduction}
In Multi-Input Multi-Output (MIMO) channels a vector $\underline{x}$ of $m_t$ signals is transmitted, a vector $\underline{y}$ of $m_r$ signals is received, and an $m_r \times m_t$ random matrix $\mathbf H$ represents the coupling of the input into the output so that the received vector is $\mathbf{\underline{y}} = \mathbf H \mathbf{\underline{x}} + \mathbf{\underline{z}}$ where $\mathbf{\underline{z}}$ is a noise vector. In this paper we consider a channel matrix $\mathbf H$ which is a sub-matrix of a Haar-distributed unitary matrix, i.e., drawn uniformly from the ensemble of all $m \times m$ unitary matrices, $m\geq m_t, m_r$.

The three classical and most well-studied random matrix ensembles are the {\emph{Gaussian}}, {\emph{Wishart}} and {\emph{Jacobi}} (also known as MANOVA) ensembles \cite{Muirhead,MehtaRandomMatrices,EdelmanRandomMatrixTheory}. The {\emph{Gaussian ensemble}} is a common model for the channel matrix $\mathbf H$ in fading wireless communication (also known as the Rayleigh model). In that case, $\mathbf H^\dag \mathbf H$ is the {\emph{Wishart ensemble}}. For the model assumed in this paper, $\mathbf H^\dag \mathbf H$ follows the {\emph{Jacobi ensemble}}. It turns out that this model is both practically useful and qualitatively different than other fading models such as the Rayleigh \cite{foschiniWirelessMIMO,TelatarCapacity,TulinoRandomMatrixTheory}, Rician \cite{OntheCapacityOfMultipleAntennaSystemsInRicianFading,CapacityOfMIMORicianChannels,DigitalCommunicationsOverFadingChannels} and Nakagami \cite{DigitalCommunicationsOverFadingChannels,Nakagami,capacityOfNakagamiChannel,RandomMatrixModelforNakagamiHoytFading}.

An important motivation to introduce such channels comes from recent developments in optical fiber communication. The expected capacity crunch in long haul optical fibers \cite{ScalingOpticalCommunications,capacityCrunch} led to proposals for ``space-division multiplexing'' (SDM) \cite{CapacityScalingThroughSpatialMultiplexing,NewGenerationOpticalInfrastructureTechnologies}, that is to have several links at the same fiber, by either multiple single-mode fiber strands within a fiber cable, multiple
cores within a multi-core fiber, or multiple modes within a multi-mode waveguide. An SDM system with $m$ parallel transmission paths per wavelength can potentially multiply the throughput of a certain link by a factor of $m$. Since $m$ can potentially be chosen very large, SDM technology is highly scalable.
Now, a significant crosstalk between the optical paths raises the need for MIMO signal processing techniques. Unfortunately, for large size MIMO  (large $m$) this is unfeasible currently in the optical rates. Assuming that faster computation will be available in the future and having in mind that replacing optical fibers to support SDM is a long and expensive procedure, a long term design is sought after. To that end and more, it was proposed to design an optical system that can support relatively large number of paths for future use, but at start to address only some of the paths. In this scenario the channel can be modeled as a sub-matrix of a larger unitary matrix, i.e., the Jacobi model is applicable.

This under-addressed channel is discussed in \cite{WinzerFoschiniOpticalMIMO} where simulations of the capacities and outage probabilities were presented. In this paper we further analyze the channel in the ergodic and non-ergodic settings, where we provide analytical expression for the capacity, outage probability and the diversity-multiplexing tradeoff.
It should be noted that in optical systems the outage probability is an important measure, required to be very low. Evidently, since the entire channel matrix is unitary, when all paths are
addressed a zero outage probability can be attained for any transmission rate.
An interesting result that comes out of this work is that there are situations,
where a partial number of pathes are addressed, yet a number of streams are guaranteed to experience zero outage.
Thus, choosing the number of addressed paths and the corresponding rate is a very critical design element that highly reflects on the system outage and performance.  A preliminary description of our work, in the context of the SDM optical channel is provided in \cite{TheUnderAddressedChannel}.

A possibly practical outcome of this work is a simple communication scheme, with channel state feedback, that achieves the highest rate possible with no outage. The scheme works even when the feedback is ``outdated'', and it allows simple decoding with no complicated MIMO signal processing, making it plausible for optical communication. The theoretical findings indicate that the no-outage promise can be attained with no feedback, yet the quest for such simple schemes is open.

While the motivation for this work comes from optical fiber communication, it should be noted that in other cases, such as in-line communication and even wireless communication, this model and the insights that follow from it can be relevant. For example, in wireless communication, it is plausible to imagine that if there were enough receive antennas capturing most, if not all, transmitted energy, the unitary assumption can be justified. In general, then, the size of the matrix $m$ with respect to $m_r$ can be viewed as a measure of the possible power loss in the medium. In a waveguide or in some in-door scenaria, the receive antennas can capture most transmitted energy, making the channel matrix almost unitary. In other cases, such as free space, much of the energy is not captured and so the channel can be modeled as a sub-matrix of a large unitary matrix. Indeed, as will be shown, when $m$ is large in comparison to $m_t,m_r$, the Jacobi model (up to a normalizing constant) approaches the Rayleigh model.

The paper is organized as follows. We start by defining the system model and presenting the channel statistics in Section \ref{sec_SystemModel}. An interesting transition threshold is revealed: when the number of addressed paths is large enough, so that $k=m_t+m_r-m>$, the statistics of the problem changes. Using this observation we give analytic expressions for the ergodic capacity in Section \ref{section_ergodicChannel}. In Section \ref{sec_NonErgodicChannel} we analyze the outage probabilities in the non-ergodic channel and show that for $k>0$ a strictly zero outage probability is obtainable for $k$ degrees of freedom. Following this finding, we present in Section \ref{sec_AchievingZeroOutageProbability} a new communication scheme which exploits a channel state feedback to achieve zero outage probability. Section \ref{sec_DMT} discuss the diversity-multiplexing tradeoff of the channel where we show an absorbing difference in the maximum diversity gain between the Rayleigh fading and Jacobi channels. Section \ref{sec_discussion} discuss the results.

\section{System Model and Channel Statistics}  \label{sec_SystemModel}
We consider a space-division multiplexing (SDM) system that supports $m$ spatial propagation paths. In tribute to optical communication, in particular multi-mode optical fibers, the initial motivation for this work, we shall refer to these links as modes. Assuming a unitary coupling among all transmission modes the overall transfer matrix $\mathbf{H}$ can be described as an $m\times m$ unitary matrix, where each entry $\mathbf{h}_{ij}$ represents the complex path gain from transmitted mode $i$ to received mode $j$. We further assume a uniformly distributed unitary coupling, that is, $\mathbf{H}$ is drawn uniformly from the ensemble of all $m\times m$ unitary matrices (Haar distributed). Considering a communication system where $m_t\leq m$ and $m_r\leq m$ modes are being addressed by the transmitter and receiver, respectively, the effective transfer matrix is a truncated version of $\mathbf{H}$. Under these conditions the channel can be described as
\begin{equation}
\mathbf{\underline{y}}=\sqrt{\rho}~\mathbf{H}_{11}\mathbf{\underline{x}}+\mathbf{\underline{z}}~, \label{channel_model_eq}
\end{equation}
where the vector $\mathbf{\underline{x}}$ containing $m_t$ complex components, represents the transmitted signal, the vector $\mathbf{\underline{y}}$
containing $m_r$ complex components, represents the received signal, and $\mathbf{\underline{z}}$ accounts for the presence of additive Gaussian noise. %The components of $\mathbf{\underline{x}}$ are constrained such that their average energy (with respect to the transmitted data) is equal to 1.
The  $m_r$ components of $\mathbf{\underline{z}}$ are statistically independent, circularly symmetric complex zero-mean Gaussian variables of unit energy ${\mathbb E}(|z_j|^2)=1$. The components of $\mathbf{\underline{x}}$ are constrained such that the average energy of each component is equal to 1, i.e., ${\mathbb E}(|x_j|^2)=1$ for all $j$ \footnote{The constant per-mode power constraint, as opposed to the constant total power constraint often used in wireless communication, is motivated by the optical fiber nonlinearity limitation. Nevertheless, the total power constraint will be considered as well when needed.}. The term $\rho\geq 0$ is proportional to the power per excited mode so that it equals to the signal-to-noise ratio in the single mode case ($m=1$). The matrix $\mathbf{H}_{11}$ is a block of size $m_r\times m_t$ within the $m\times m$ random unitary matrix $\mathbf H$
%
%\begin{equation}
\begin{equation}
\mathbf{H}=\begin{bmatrix}
\mathbf{H}_{11} &\mathbf{H}_{12}\\
\mathbf{H}_{21} &\mathbf{H}_{22}\\
\end{bmatrix}~.
\label{matrix H}
\end{equation}

As a first stage in our analysis we establish the relation between the transfer matrix $\mathbf{H}_{11}$ and the Jacobi ensemble of random matrices \cite{Muirhead,MehtaRandomMatrices,EdelmanRandomMatrixTheory}. Limiting our discussion to complex matrices we state the following definitions:

\begin{defn} [Gaussian ensemble]
$\mathcal{G}(m,n)$ is $m\times n$ matrix of i.i.d complex entries distributed as $\mathcal{CN}(0,1)$.
\end{defn}
\begin{defn} [Wishart ensemble]
$\mathcal{W}(m,n)$, where $m\geq n$, is $n\times n$ Hermitian matrix which can be constructed as $A^\dag A$, where $A$ is $\mathcal{G}(m,n)$.
\end{defn}
\begin{defn} [Jacobi ensemble]
$\mathcal{J}(m_1,m_2,n)$, where $m_1,m_2\geq n$, is $n\times n$ Hermitian matrix which can be constructed as $A(A+B)^{-1}$, where $A$ and $B$ are $\mathcal{W}(m_1,n)$ and $\mathcal{W}(m_2,n)$, respectively.
\end{defn}

The first two ensembles relate to wireless communication \cite{TulinoRandomMatrixTheory}. We claim here that the third classical ensemble, the Jacobi ensemble, is relevant to this channel model by relating its eigenvalues to the singular values of $\mathbf{H}_{11}$. To that end we quote the well-known \cite{Muirhead} joint probability density function (PDF) of the ordered eigenvalues $0\leq \lambda_1\leq \ldots \leq \lambda_{n}\leq 1$ of the Jacobi ensemble $\mathcal{J}(m_1,m_2,n)$
\begin{equation}
f(\lambda_1^{n})=K^{-1}_{m_1,m_2,n}\prod_{i=1}^{n} \lambda_i ^{m_1-n}(1-\lambda_i)^{m_2-n}\prod_{i<j}(\lambda_i-\lambda_j)^2~, \label{JacobiPDF}
\end{equation}
where $K_{m_1,m_2,n}$ is a normalizing constant. We say that $n$ variables follow the law of the Jacobi ensemble $\mathcal{J}(m_1,m_2,n)$ if their joint distribution follows \eqref{JacobiPDF}.

We shall now present the explicit distribution of the channel's singular values by distinguishing between the following two cases:

\subsection{Case \Rmnum{1} - $m_t+m_r\leq m$} \label{subsection_firstCase_statistics}
In \cite[Theorem 1.5]{EdelmanCSdecomposition} it was shown that for $m_t,~m_r$ satisfying the conditions $m_t\leq m_r$ and $m_t+m_r\leq m$, the eigenvalues of $\mathbf{H}_{11}^\dag \mathbf{H}_{11}$ have the same distribution as the eigenvalues of the Jacobi ensemble $\mathcal{J}(m_r,m-m_r,m_t)$. For $m_t,~m_r$ satisfying $m_t> m_r$ and $m_t+m_r\leq m$, since $\mathbf{H}^\dag$ share the same distribution with $\mathbf{H}$, the eigenvalues of $\mathbf{H}_{11} \mathbf{H}_{11}^\dag$ follow the law of the Jacobi ensemble $\mathcal{J}(m_t,m-m_t,m_r)$. Combining these two results, we can say that the squared non-zero singular values of $\mathbf{H}_{11}$ have the same distribution as the eigenvalues of the Jacobi ensemble $\mathcal{J}(m_{\text{max}},m-m_{\text{max}},m_{\text{min}})$, where here and throughout this paper we denote $m_{\text{max}}=\max\{m_t,m_r\}$ and $m_{\text{min}}=\min\{m_t,m_r\}$.

%EXPLAIN ON THE DISTRIBUTION OF GAUSSIAN MATRICES - Ronen.\\
%EXPLAIN ON THE DISTRIBUTION - GRAPHS ON THE EXTREME VALUE - Ronen.

\subsection{Case \Rmnum{2} - $m_t+m_r> m$}
When the sum of transmit and receive modes, $m_t+m_r$, is larger than the total available modes, $m$, the statistics of the singular values change. Having in mind that the columns of $\mathbf{H}$ are orthonormal, one can think of $m_t+m_r>m$ as a transition threshold in which the size of $\mathbf{H}_{11}$ is large enough with respect to $m$ to change the singularity statistics. The following Lemma provides the joint distribution of the singular values of $\mathbf{H}_{11}$, showing that for any realization of $\mathbf{H}_{11}$ there are $m_t+m_r-m$ singular values which are $1$.

\begin{lem} \label{lem1}
Suppose $\textnormal H$
is an $m\times m$ unitary matrix, divided into blocks as in (\ref{matrix H}), where
$\textnormal{H}_{11}$ is an $m_r\times m_t$ block with $m_t+m_r> m$. Then $m_t+m_r-m$ eigenvalues of $\textnormal{H}_{11}^\dag \textnormal{H}_{11}$ are 1, $m_t-m_{\text{min}}$ are 0, and $m-m_{\text{max}}$ are equal to the non-zero eigenvalues of $\textnormal{H}_{22} \textnormal{H}_{22}^\dag$; thus, if $\mathbf{H}$ is Haar distributed these $m-m_{\text{max}}$ eigenvalues follow the law of the Jacobi ensemble
$\mathcal{J}(m-m_{\text{min}},m_{\text{min}},m-m_{\text{max}})$.
\end{lem}
\begin{proof}
Since $\text H$ unitary we can write
\begin{equation}
\text{H}_{11}^\dag \text{H}_{11}+\text{H}_{21}^\dag \text{H}_{21}=\text{I}_{m_t} \label{lem1_eq1}
\end{equation}
and
\begin{equation}
\text{H}_{21} \text{H}_{21}^\dag+\text{H}_{22} \text{H}_{22}^\dag=\text{I}_{m-m_r}~.\label{lem1_eq4}
\end{equation}
Let $\{\lambda_i^{(11)}\}_{i=1}^{m_t}$ and $\{\lambda_i^{(21)}\}_{i=1}^{m_t}$ be the eigenvalues of $\text{H}_{11}^\dag \text{H}_{11}$ and $\text{H}_{21}^\dag \text{H}_{21}$, respectively. From \eqref{lem1_eq1} we can write
\begin{equation}
\lambda^{(11)}_i=1-\lambda^{(21)}_i \qquad \forall~i=1,\ldots,m_t~. \label{lem1_eq2}
\end{equation}
Since $\text{H}_{21}$ is a block of size $(m-m_r)\times m_t$ where $m-m_r<m_t$, $\text{H}_{21}^\dag \text{H}_{21}$ has (at least) $m_t+m_r-m$ zero eigenvalues. Following \eqref{lem1_eq2}, $\text{H}_{11}^\dag \text{H}_{11}$ has $m_t+m_r-m$ eigenvalues which are 1. Now, let $\{\tilde\lambda_i^{(21)}\}_{i=1}^{m-m_r}$ and $\{\tilde\lambda_i^{(22)}\}_{i=1}^{m-m_r}$ be the eigenvalues of $\text{H}_{21} \text{H}_{21}^\dag$ and $\text{H}_{22} \text{H}_{22}^\dag$, respectively. From \eqref{lem1_eq4} we can write
\begin{equation}
\tilde\lambda^{(21)}_i=1-\tilde{\lambda}^{(22)}_i \qquad \forall~i=1,\ldots,m-m_r~. \label{lem1_eq3}
\end{equation}
Since for any matrix $A$, $A^\dag A$ and $A A^\dag$ share the same non-zero eigenvalues we can combine \eqref{lem1_eq2} and \eqref{lem1_eq3} to conclude that the additional $m-m_{r}$ eigenvalues of $\text{H}_{11}^\dag \text{H}_{11}$ are equal to the $m-m_{r}$ eigenvalues of $\text{H}_{22} \text{H}_{22}^\dag$. Note that $m_t-m_{\text{min}}$ of them are 0.
Since the above arguments hold for any unitary matrix, and since $\text{H}_{22}$ is a block of size $(m-m_r)\times (m-m_t)$, when $\mathbf{H}$ is Haar distributed the results of subsection \ref{subsection_firstCase_statistics} can be applied, which completes the proof.\\
\end{proof}

Lemma \ref{lem1} reveals an interesting algebraic phenomenon: $k=\max\{m_t+m_r-m,0\}$ singular values of $\mathbf{H}_{11}$ are 1 for any realization of $\mathbf{H}$. This provides some powerful results in the context of Jacobi fading channels. For example, the channel's power $\|\mathbf{H}_{11}\|_F^2$, where $\|A\|_F$ denotes the Frobenius norm of $A$, is guaranteed to be at least $k$. Furthermore, $\mathbf{H}_{11}$ always comprises an unfaded $k$-dimensional subspace. In what follows we show that this implies a lower bound on the ergodic capacity, an achievable zero outage probability and an ``unbounded'' diversity gain for certain rates.

%This phenomenon also has a geometric interpretation: $H_{11}$ and $H_{21}$ form a set of $m_t$ orthonormal vectors, i.e., $H_{21}$ completes $H_{11}$ into a vector space $\mathcal V$ with an orthonormal basis $U$. $H_{21}$ is a block of size $(m-m_r)\times m_t$, therefore if $m-m_r<m_t$, $H_{21}$ is an $(m-m_r)$-dimensional subspace of $\mathcal V$ with basis $D U$, where $D$ is diagonal with only $m-m_r$ non-zero elements. Thus the basis of $H_{11}$ is $(\text{I}_{m_t}-D)U$, that is, comprises at least $m_t-(m-m_r)$ vectors of norm 1.

\section{The Ergodic Case} \label{section_ergodicChannel}
In the ergodic scenario the channel is assumed to be rapidly changing so that the transmitted signal samples the entire channel statistics. We further assume that the channel realization at each symbol time is known only at the receiver end. It is well known \cite{TelatarCapacity} that the channel capacity in that case is achieved by taking $\mathbf{\underline{x}}$ to be a vector of circularly symmetric complex zero-mean Gaussian components; and is given by
\begin{equation}
C({\text{\small{$m_t,m_r,m;\rho$}}})=\max_{\substack{Q:~Q\succeq 0\\Q_{ii}\leq 1~\forall~i=1,\ldots,m_t}}\mathbb{E}[\log \det(\text{I}_{m_r}+\rho \mathbf{H}_{11}Q\mathbf{H}_{11}^\dag)]~, \label{eqNew1}
\end{equation}
where the maximization is over all covariance matrices of $\mathbf{\underline{x}}$, $Q$, that satisfy the power constraints. Now, the capacity in \eqref{eqNew1} also satisfies
\begin{align}
C({\text{\small{$m_t,m_r,m;\rho$}}})&\leq \max_{\substack{Q:~Q\succeq 0\\trace(Q)\leq m_t}}\mathbb{E}[\log \det(\text{I}_{m_r}+\rho  \mathbf{H}_{11}Q\mathbf{H}_{11}^\dag)]~, \label{eqThmCap1}
\end{align}
where it is well known \cite[Theorem 1]{TelatarCapacity} that if the distribution of $\mathbf{H}_{11}$ is invariant under unitary permutations, $Q=\text{I}_{m_t}$ is the optimal choice for \eqref{eqThmCap1}. Since $\mathbf{H}$ is Haar-distributed, that is invariant under unitary permutations, also $\mathbf{H}_{11}$ is invariant under unitary permutations. Thus $Q=\text{I}_{m_t}$ is the optimal choice for \eqref{eqNew1} and by using the following equation
\begin{equation*}
\det(\text{I}_{m_r}+\rho  \text{H}_{11}\text{H}_{11}^\dag)=\log \det(\text{I}_{m_t}+\rho  \text{H}_{11}^\dag\text{H}_{11}),
\end{equation*}
we can conclude that the ergodic capacity is given by
\begin{equation}
C({\text{\small{$m_t,m_r,m;\rho$}}})=\mathbb{E}[\log \det(\text{I}_{m_t}+\rho \mathbf{H}_{11}^\dag\mathbf{H}_{11})]~. \label{eq_ergodicCapacity}
\end{equation}

%Before we continue we note that the ergodic capacity is symmetric in $m_t$ and $m_r$ since we applied a constant {\emph per-mode} power constraint.%; since the transmitter has no information about the channel realization, the direction of the transmitted signal can be chosen arbitrary.

\subsection{Case \Rmnum{1} - $m_t+m_r\leq m$} \label{subsection_ergodicCap1}
The following theorem gives an analytical expression to the ergodic capacity for cases where $m_t+m_r\leq m$. Using the joint distribution of the eigenvalues of the Jacobi ensemble we associate the ergodic capacity with the Jacobi polynomials \cite[8.96]{tableOfIntegrals}.

\begin{thm} \label{ergodicCap_int}
The ergodic capacity of the channel defined in \eqref{channel_model_eq} with $m_t,~m_r$ satisfying $m_t+m_r\leq m$, reads
\begin{align}
C({\text{\small{$m_t,m_r$}}}&{\text{\small{$,m;\rho$}}})=\int_0^1 \lambda^{\alpha}(1-\lambda)^{\beta}\log(1+\lambda \rho) \sum_{k=0}^{m_{\text{min}}-1} b_{k,\alpha,\beta}^{-1}[P^{(\alpha,\beta)}_{k}(1-2\lambda)]^2d\lambda
\end{align}
where $P^{(\alpha,\beta)}_{k}(x)$ are the Jacobi polynomials
\begin{align}
P^{(\alpha,\beta)}_{k}(x)=\tfrac{(-1)^{k}}{2^kk!}&(1-x)^{-\alpha}(1+x)^{-\beta}\frac{d^k}{dx^k}\big[(1-x)^{k+\alpha}(1+x)^{k+\beta}\big]~, \label{JacobiPol}
\end{align}
the coefficients $b_{k,\alpha,\beta}$ are given by
\begin{equation*}
b_{k,\alpha,\beta}=\frac{1}{2k+\alpha+\beta+1}{2k+\alpha+\beta \choose k} {2k+\alpha+\beta \choose k+\alpha}^{-1}~,
\end{equation*}
and $\alpha=\abs{m_r-m_t}$, $\beta=m-m_t-m_r$.
\end{thm}

\begin{proof}
See Appendix \ref{appendix_ergodicCapcaty}.
\end{proof}

\begin{figure}[t]
\begin{subfigure}[t]{0.5\textwidth}
\includegraphics[height=0.76\textwidth,width=1\textwidth]{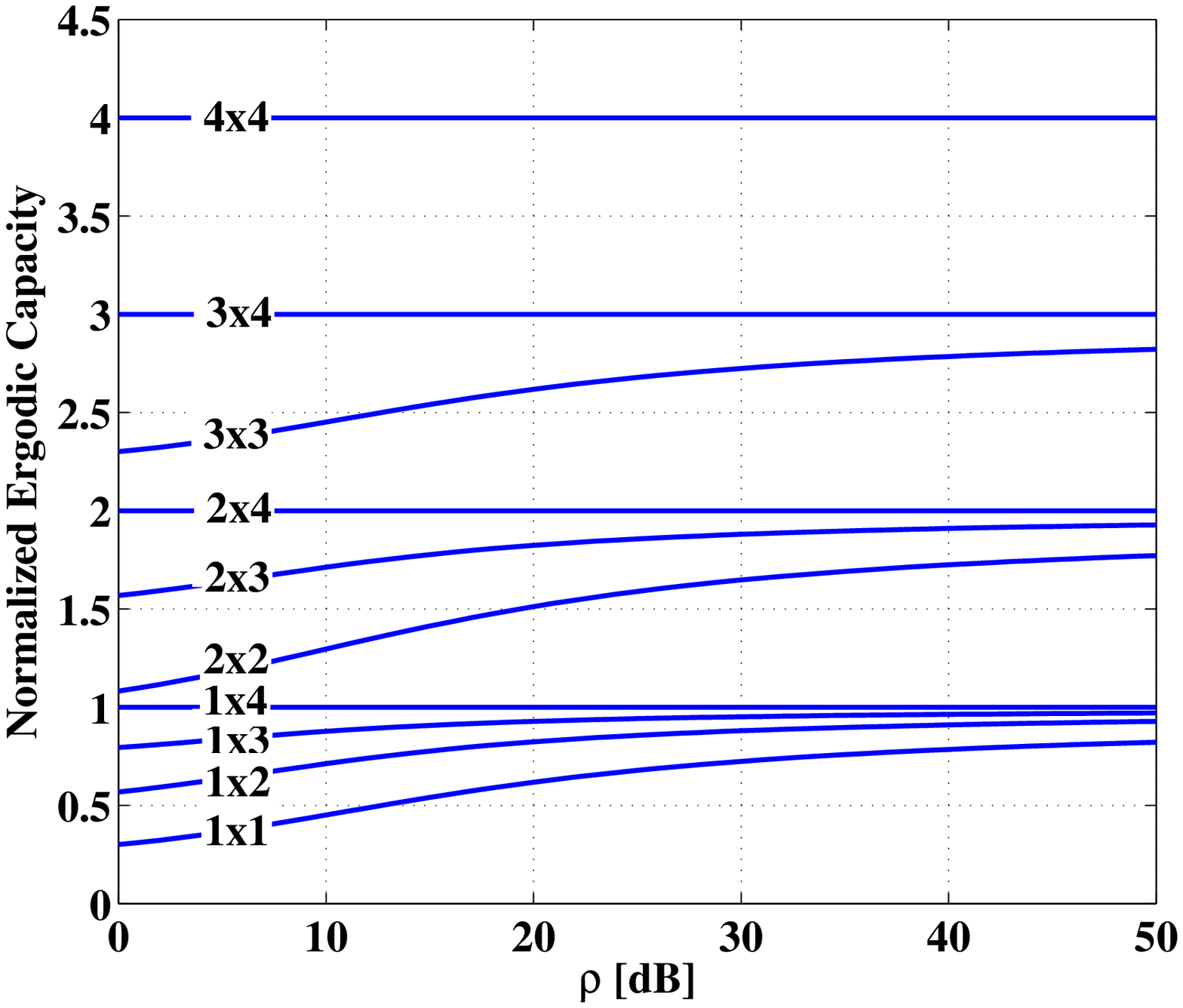}
\caption{} \label{ergodicCapacity_n6_1}
\end{subfigure}\hspace{0.2cm}
\begin{subfigure}[t]{0.5\textwidth}
\includegraphics[height=0.78\textwidth,width=1.05\textwidth]{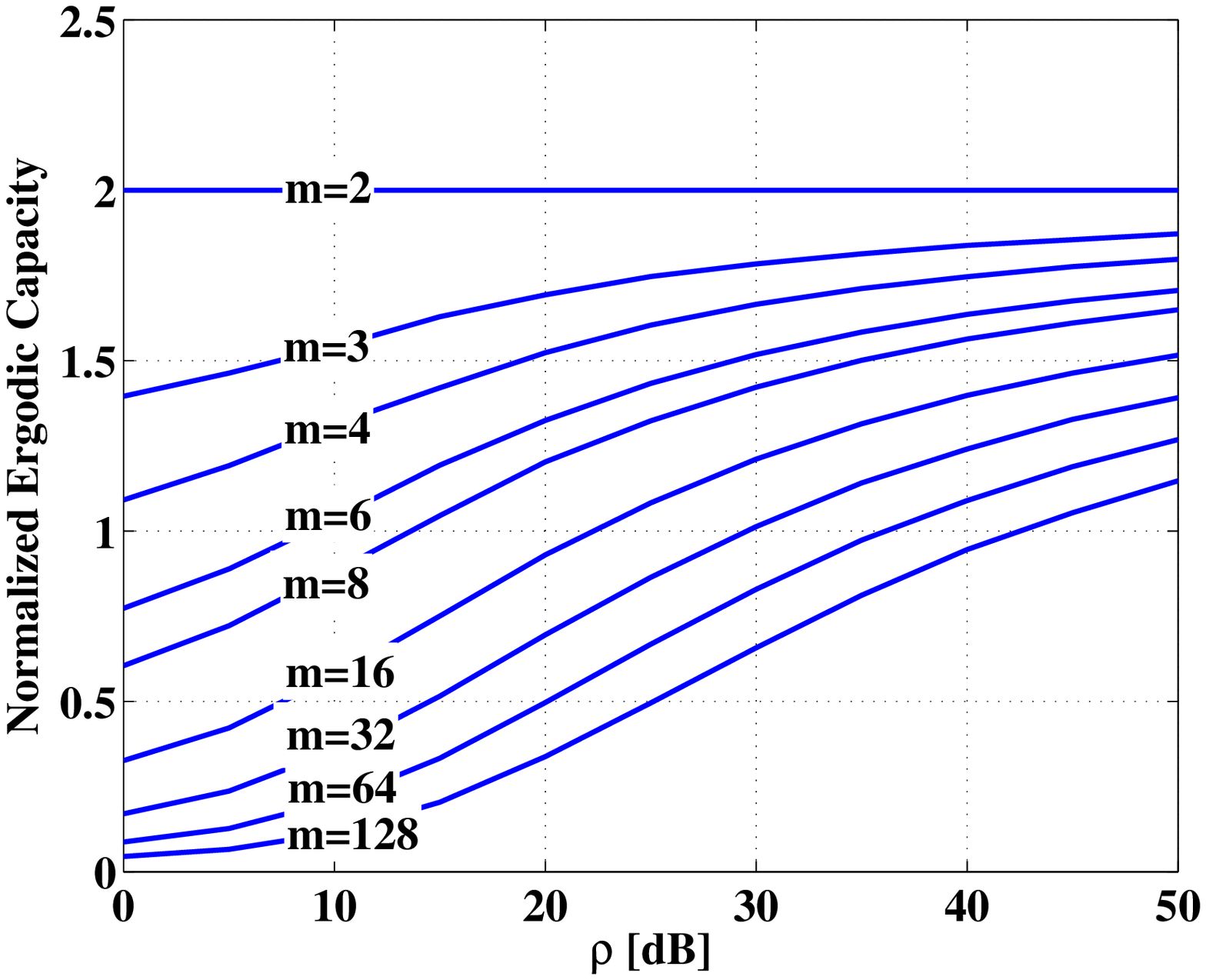}
\caption{} \label{ergodicCapacity_n6_2}
\end{subfigure}
\caption{The ergodic capacity, normalized by $C({\text{\small{$1,1,1;\rho$}}})=\log(1+\rho)$, as a function of $\rho$. In (\subref{ergodicCapacity_n6_1}) the number of supported modes is fixed $m=4$, various numbers of transmit$\times$receive modes; in (\subref{ergodicCapacity_n6_2}) the number of addressed modes is fixed $m_t=m_r=2$, various values of supported modes $m$.} \label{ergodicCapacity_n6}
\end{figure}

\subsection{Case \Rmnum{2} - $m_t+m_r> m$}
Applying Lemma \ref{lem1} to the channel capacity given in \eqref{eq_ergodicCapacity} readily results in the following theorem.

\begin{thm} \label{thm_ergodicCap2case}
The ergodic capacity of the channel defined in \eqref{channel_model_eq} with $m_t,~m_r$ satisfying $m_t+m_r> m$, reads
\begin{align}
C({\text{\small{$m_t,m_r,m;\rho$}}})=&(m_t+m_r-m)  C({\text{\small{$1,1,1;\rho$}}})+ C({\text{\small{$m-m_r,m-m_t,m;\rho$}}}) ~, \label{500}
\end{align}
where $C({\text{\small{$1,1,1;\rho$}}})$ is the SISO channel capacity $\log(1+\rho)$.
\end{thm}
\begin{proof}
According to \eqref{eq_ergodicCapacity} the ergodic capacity satisfies
\begin{align}
C({\text{\small{$m_t,m_r,m;\rho$}}})&=\mathbb{E}[\log \det(\text{I}_{m_t}+\rho  \mathbf{H}_{11}^\dag\mathbf{H}_{11})]\\
&=\mathbb{E}[\sum_{i=1}^{m_t}\log (1+\rho \lambda_i)]~, \label{thm2_eq1}
\end{align}
where $\{\lambda_i\}_{i=1}^{m_t}$ are the eigenvalues of $\mathbf{H}_{11}^\dag\mathbf{H}_{11}$. According to Lemma \ref{lem1}, $m_t+m_r-m$ eigenvalues are $1$ and the rest are equal to the $m-m_r$ eigenvalues of $\mathbf{H}_{22} \mathbf{H}_{22}^\dag$. Applying that into \eqref{thm2_eq1} results
\begin{equation}
C({\text{\small{$m_t,m_r,m;\rho$}}})=(m-m_r-m_t)\log (1+\rho)+\mathbb{E}[\log \det(\text{I}_{m-m_r}+\rho \mathbf{H}_{22} \mathbf{H}_{22}^\dag)\big)]~.
\end{equation}
\end{proof}

\vspace{0.5cm}

Note that the second term on the right-hand-side of (\ref{500}), $C({\text{\small{$m-m_r,m-m_t,m;\rho$}}})$, is given by Theorem \ref{ergodicCap_int} and reduces to 0 when $m_t$, or $m_r$ is equal to $m$. Thus, (\ref{500}) suggests that for systems with $k=m_t+m_r-m>0$, the ergodic capacity is the sum capacities of $k$ unfaded SISO capacities and a Jacobi MIMO channel with $m-m_r$ transmit modes and $m-m_t$ receive modes. Fig. \ref{ergodicCapacity_n6_1} depicts the ergodic capacity as a function of $\rho$ for $m=4$ and various combinations of $m_t,~m_r$ (note that the ergodic capacity, in our case, is symmetric in $m_t,~m_r$; thus all combinations are plotted). As is evident from the figure, a capacity equivalent to $k$ SISO channels is guaranteed in all cases. In Fig. \ref{ergodicCapacity_n6_2} the ergodic capacities for $m_t=m_r=2$ and various values of supported modes are plotted. Note that as $m$ increases, the power loss increases and the ergodic capacity becomes smaller. Unlike the common practice of expressing the capacity in terms of the received SNR, here the capacities are presented as a function of $\rho$. This normalizes the capacity expression to reflect the capacity loss due to power loss including power leaked into the unobserved modes. In particular, this presentation enables to examine the total effect (capacity loss) of increasing $m$. See further discussion in section \ref{section_Rayleigh}.

\section{The Non-Ergodic Case} \label{sec_NonErgodicChannel}
In the non-ergodic scenario the channel matrix is drawn randomly but rather assumed to be constant within the entire transmission period of each code-frame. The figure of merit in the non-ergodic case is the {\emph{outage probability}} defined as the probability that the mutual information induced by the channel realization is lower than the rate $R$ at which the link is chosen to operate. Note that we assume that the channel instantiation is unknown at the transmitter, thus it can not adapt the transmission rate. However, the channel is assumed to be known at the receiver end. By taking an input vector of circularly symmetric complex zero-mean Gaussian variables with covariance matrix $Q$ the mutual information is maximized and the outage probability can be expressed as
\begin{align}
&{P}_{out}\text{\small{$(m_t,m_r,m;R)$}}=\inf_{\substack{Q:~Q\succeq 0}}Pr\big[\log \det(\text{I}_{m_r}+\rho  \mathbf{H}_{11}Q\mathbf{H}_{11}^\dag)<R\big],
\end{align}
where the minimization is over all covariance matrices $Q$ satisfying the power constraints. Since the statistics of $\mathbf{H}_{11}$ is invariant under unitary permutations, the optimal choice of $Q$, when applying {\emph {constant per-mode power constraint}}, is simply the identity matrix. We note that when imposing {\emph {total power constraint}}, the optimal choice of $Q$ may depend on $R$ and $\rho$ and in general is unknown, even for the Rayleigh channel. Nevertheless, when $\rho\gg 1$ the identity matrix is approximately the optimal $Q$ (see section \ref{sec_DMT}).  Thus, in the following we make the simplified assumption that the transmitted covariance matrix is the commonly used choice $Q=\text{I}_{m_t}$. %We denote the outage probability in that case by $P_{out}\text{\small{$(m_t,m_r,m;R)$}}$.

Now, let the transmission rate be $R=r\log(1+\rho)$ (bps/Hz) and let $\underline{\lambda}=\{\lambda_i\}_{i=1}^{m_{\text{min}}}$ be the ordered non-zeros eigenvalues of $\mathbf{H}_{11}^\dag\mathbf{H}_{11}$; we can write
\begin{align}
P_{out}\text{\small{$(m_t,m_r,m;r\log(1+\rho))$}}&=Pr\big[\log \det(\text{I}_{m_t}+\rho \mathbf{H}_{11}^\dag\mathbf{H}_{11})<R\big]\\
&= Pr\big[\prod_{i=1}^{m_{\text{min}}} (1+\rho \lambda_i)<(1+\rho)^r\big]~, \label{outagePr_1}
\end{align}
and evaluate this expression by applying the statistics of $\underline{\lambda}$.

\subsection{Case \Rmnum{1} - $m_t+m_r\leq m$} \label{subsection_outagePr1}
Using \eqref{JacobiPDF} we can apply the joint distribution of $\underline{\lambda}$ into \eqref{outagePr_1} to get
\begin{align}
P_{out}&\text{\small{$(m_t,m_r,m;r\log(1+\rho))$}}= K^{-1}_{m_t,m_r,m}\int_{\mathcal{B}}\prod_{i=1}^{m_{\text{min}}} \lambda_i ^{\abs{m_r-m_t}}(1-\lambda_i)^{m-m_r-m_t}\prod_{i<j}(\lambda_i-\lambda_j)^2d \mathbf{\underline{\lambda}}~,\label{ouragePr_fin}
\end{align}
where $K_{m_t,m_r,m}$ is a normalizing factor and $\mathcal{B}$ describes the outage event
\begin{align*}
\mathcal{B}=\bigg\{\underline{\lambda} :~\prod_{i=1}^{m_{\text{min}}} (1+\rho \lambda_i)<(1+\rho)^r\bigg\}~.
\end{align*}
This gives an analytical expression to the outage probability. See Fig. \ref{outagePr_n4} and the example below.

\begin{figure}[t]
\center
\psfrag{X}{\vspace{-0.3cm}\textbf{\footnotesize{$m_r/m$}}}
\psfrag{Y}{\textbf{\footnotesize{$\rho_{\text{norm}}~[\text{dB}]$}}}
\epsfig{file=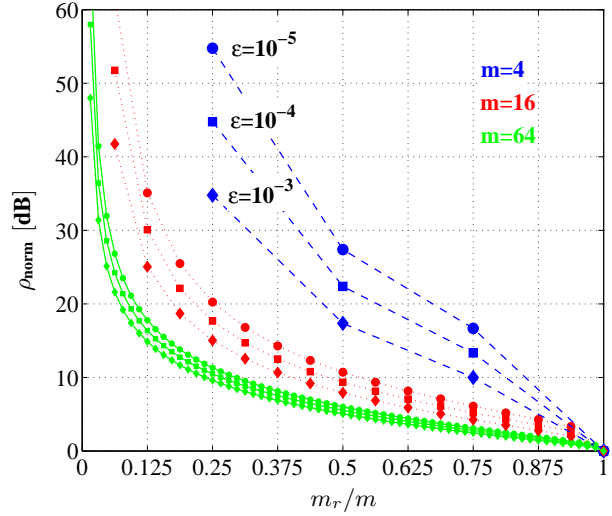,height=0.43\textwidth,width=0.55\textwidth,clip=}
%\caption{$l(\epsilon,m_r,m-m_r)$ as a function of $m_r$ for $m-m_r=1,10,100,1000$. Curves are drawn for outage probabilities $\epsilon=10^{-2}, 10^{-3},10^{-4},10^{-5}$.}
\caption{$\rho_{\text{norm}}$ as a function of $m_r/m$ for $m=4,16,64$ (blue dashed, red dotted, green solid). Curves are drawn for outage probabilities $\epsilon= 10^{-5},10^{-4},10^{-3}$ (circle, square, diamond).} \label{fig_outage_pr_mr_vs_m}
\end{figure}

\begin{example}
Suppose $m_t=1$ and $m_r,~m$ satisfy $m\geq 1+m_r$. In that case the outage probability is given by
\begin{equation}
P_{out}\text{\small{$(1,m_r,m;R)$}}=K^{-1}_{1,m_r,m}\int_0^{(2^R-1)/\rho}\lambda^{m_r-1}(1-\lambda)^{m-m_r-1}d\lambda~.
\end{equation}
Thus, we can write
\begin{equation}
P_{out}\text{\small{$(1,m_r,m;R)$}}=\frac{B((2^R-1)/\rho;m_r,m-m_r)}{B(1;m_r,m-m_r)}~,
\end{equation}
where $B(x;a,b)$ is the incomplete beta function. Hence, to support an outage probability smaller than $\epsilon$, $R$ and $\rho$ have to satisfy
\[\frac{\rho}{2^R-1}\geq \rho_{\text{norm}}=1/B^{-1}(\epsilon B(1;m_r,m-m_r);m_r,m-m_r)~,\]
where $B^{-1}(x;a,b)$ is the inverse function of $B(x;a,b)$. $\rho_{\text{norm}}$ is the {\emph{normalized}} signal-to-noise ratio at the transmitter, is proportional to the received {\emph{normalized}} signal-to-noise ratio, and essentially measures the minimal additional power required to support a target rate $R$ with outage probability smaller than $\epsilon$ (additional power over the minimal required in SISO unfading channel ($m=m_r$)). As $\rho_{\text{norm}}$ is smaller one can afford higher data rate or smaller $\rho$ (smaller transmission power).
%In Fig. \ref{fig_outage_pr_mr_vs_m} we plot $l(\epsilon,m_r,m-m_r)$ as a function of $m_r$ for various numbers of unaddressed receive modes $m-m_r$ and desired outage probabilities $\epsilon$. For fixed $m_r$ and $m-m_r$, $l(\epsilon,m_r,m-m_r)$ decreases as $\epsilon$ decreases (since more power or lower data rate are needed to achieve smaller outage probability). This is also true when $\epsilon$ and $m_r$ are fixed and $m-m_r$ increases (since more power is lost in the unaddressed receive modes). For fixed $\epsilon$ and $m-m_r$, as $m_r$ is larger, $l(\epsilon,m_r,m-m_r)$ is higher (since the diversity at the receiver is higher, see Section \ref{sec_DMT}). Note that as $m_r$ is larger, the number of unaddressed receive modes $m-m_r$ turns neglectable and $l(\epsilon,m_r,m-m_r)$ approaches to 1.
\newline In Fig. \ref{fig_outage_pr_mr_vs_m} we plot $\rho_{\text{norm}}$ as a function of $m_r/m$ for various numbers of available modes $m$ and desired outage probabilities $\epsilon$. For fixed $m$ and $m_r/m$, $\rho_{\text{norm}}$ increases as $\epsilon$ decreases (since more power or lower data rate are needed to achieve smaller outage probability). For fixed $\epsilon$ and $m$, $\rho_{\text{norm}}$ decreases as $m_r/m$ increases (since more modes are addressed by the receiver, therefore the power loss decreases). This is also true as $m$ increases while $\epsilon$ and $m_r/m$ are fixed (since the diversity at the receiver increases, see Section \ref{sec_DMT}). Note that for $m_r/m=1$ there is no power loss and we get $\rho_{\text{norm}}=1$, that is, the minimal transmission power required to support the rate $R$, for any $\epsilon$, is $\rho =2^R-1$.
\end{example}

\subsection{Case \Rmnum{2} - $m_t+m_r> m$}
Applying Lemma \ref{lem1} into \eqref{outagePr_1} results the following.

\begin{thm} \label{thm_outagePr2case}
The outage probability of the channel defined in \eqref{channel_model_eq}, with $m_t,~m_r$ satisfying $m_t+m_r>m$, satisfies
\begin{align}
P&_{out}\text{\small{$(m_t,m_r,m;r\log(1+\rho))$}}=P_{out}\text{\small{$(m-m_r,m-m_t,m;\tilde{r}\log(1+\rho))$}}~,\label{secondCaseOutPr}
\end{align}
where $\tilde{r}$ is the larger between $r-(m_t+m_r-m)$ and 0.
\end{thm}
\begin{proof}
According to \eqref{outagePr_1}, the outage probability is given by
\begin{equation}
P_{out}\text{\small{$(m_t,m_r,m;r\log(1+\rho))$}}= Pr\big[\prod_{i=1}^{m_t} (1+\rho \lambda_i)<(1+\rho)^r\big]~,
\end{equation}
where $\{\lambda_i\}_{i=1}^{m_t}$ are the eigenvalues of $\mathbf{H}_{11}^\dag\mathbf{H}_{11}$. By applying Lemma \ref{lem1} we get
\begin{align}
P_{out}\text{\small{$(m_t,m_r,m;r\log(1+\rho))$}}&= Pr\big[\prod_{i=1}^{m-m_r} (1+\rho \tilde{\lambda}_i)<(1+\rho)^{r-(m_t+m_r-m)}\big]~,
\end{align}
where  $\{\tilde{\lambda}_i\}_{i=1}^{m-m_r}$ are the eigenvalues of $\mathbf{H}_{22}\mathbf{H}_{22}^\dag$. When $\tilde r=r-(m_t+m_r-m)<0$ we get $P_{out}\text{\small{$(m_t,m_r,m;r\log(1+\rho))$}}=0$.
\end{proof}

\vspace{0.5cm}

\begin{figure}[t]
\begin{subfigure}[t]{0.5\textwidth}
\includegraphics[height=0.76\textwidth,width=1\textwidth]{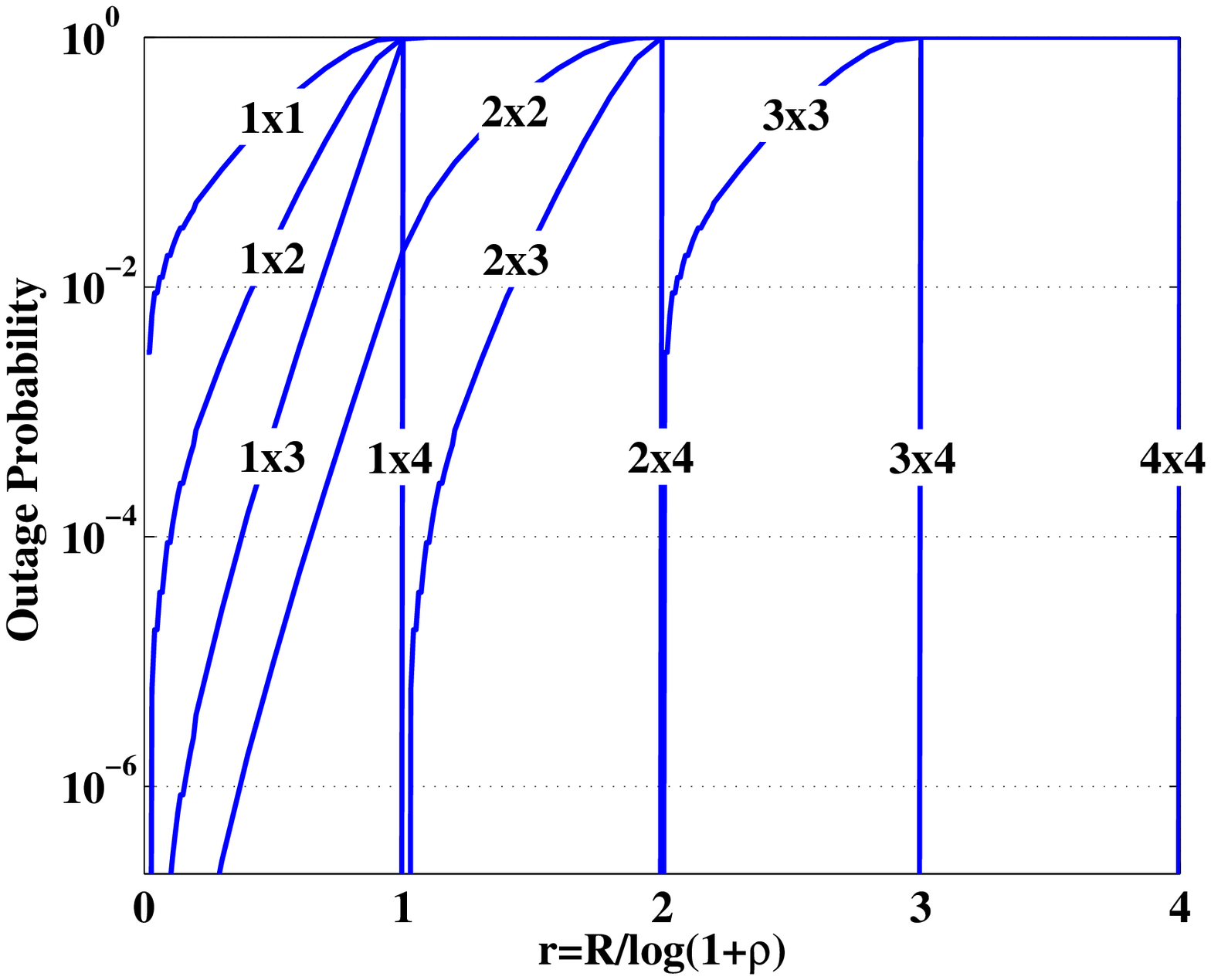}
\caption{} \label{outagePr_n4_1}
\end{subfigure}\hspace{0.5cm}
\begin{subfigure}[t]{0.5\textwidth}
\includegraphics[height=0.75\textwidth,width=0.99\textwidth]{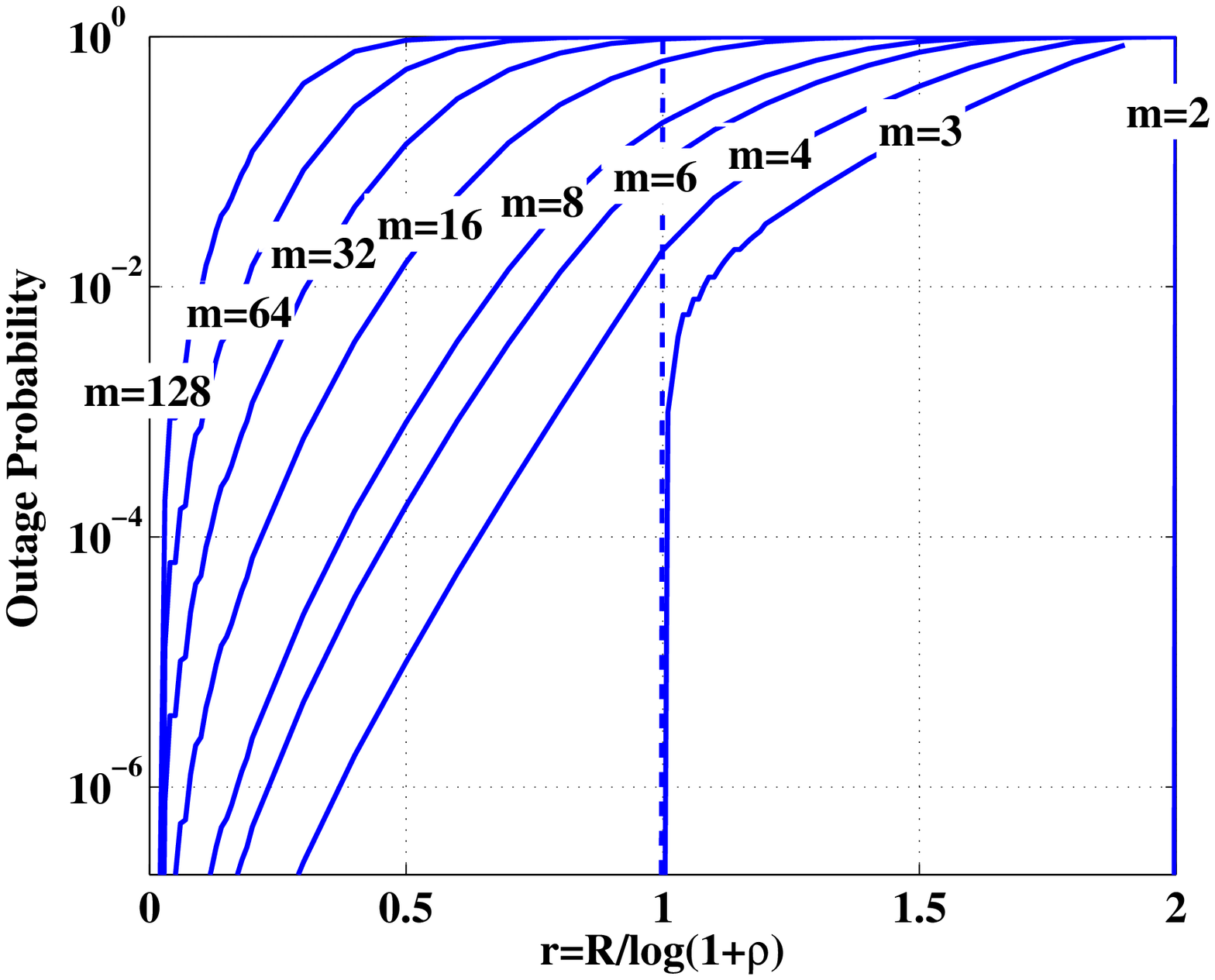}
\caption{} \label{outagePr_n4_2}
\end{subfigure}
\caption{Outage probability vs. normalized rate for $\rho=20$dB. In (\subref{outagePr_n4_1}) the number of supported modes is fixed $m=4$, various numbers of transmit$\times$receive modes; in (\subref{outagePr_n4_2}) the number of addressed modes is fixed $m_t=m_r=2$, various values of supported modes $m$.} \label{outagePr_n4}
\end{figure}

Note that the right-hand-side drops to 0, when $m_r$, or $m_t$ equals $m$. Most importantly, when $r<m_t+m_r-m$, $\tilde r=0$, implying that for such rates zero outage probability is achievable. In addition, when $r\geq m_t+m_r-m>0$, Eq. \eqref{secondCaseOutPr} implies that the outage probability is identical to that of a channel with $m-m_r$ modes addressed by the transmitter and $m-m_t$ modes addressed by the receiver, which is designed to support a transmission rate equivalent to $\tilde r$ single-mode channels. Thus the right-hand-side of \eqref{secondCaseOutPr} applies to Eq. \eqref{ouragePr_fin}. In Fig. \ref{outagePr_n4_1} we show an exemplary calculation of the outage probability. These curves, obtained from our analysis were plotted in the same form as the numerical results reported in \cite{WinzerFoschiniOpticalMIMO}. Note how the outage probability abruptly drops to 0 whenever $r$ becomes smaller than $m_t+m_r-m$. Also note that the outage probability is symmetric in $m_t,~m_r$ since we applied a constant {\emph {per-mode}} power constraint; thus all combinations of $m_t,~m_r$ are plotted in Fig. \ref{outagePr_n4_1}. In Fig. \ref{outagePr_n4_2} outage probability curves are plotted for $m_t=m_r=2$ and various values of supported modes, $m$. Note that as $m$ is larger, more power is lost in the unaddressed modes, therefore, as evident from the figure, the outage probability increases.

\section{Achieving The No-Outage Promise} \label{sec_AchievingZeroOutageProbability}
In the previous section we saw that for systems satisfying $k=m_t+m_r-m>0$, a zero outage probability is achievable for any transmission rate below $R=k\log(1+\rho)$. In this section we present a new communication scheme that achieves this promise with a transmission rate arbitrarily close to $R=k\log(1+\rho)$. Using simple manipulations, the scheme exploits a (delayed) channel state information (CSI) feedback to transform the channel into $k$ independent SISO channels, supporting $k$ streams (degrees of freedom) with zero outage probability. %Furthermore, our scheme removes the need for MIMO decoding and allows the use of simple SISO channel decoders.

Let \[\mathbf{H}^{(i)}=\begin{bmatrix}
\mathbf{H}^{(i)}_{11} &\mathbf{H}^{(i)}_{12}\\
\mathbf{H}^{(i)}_{21} &\mathbf{H}^{(i)}_{22}\\
\end{bmatrix}\]
be the unitary matrix realization at channel use $i$ and let
\[{\mathbf{\underline{y}}}^{(i)}= \sqrt{\rho}\mathbf{H}^{(i)}_{11}\mathbf{\underline{x}}^{(i)}+{\mathbf{\underline{z}}}^{(i)} \]
be the received signal. We assume a perfect knowledge of $\mathbf{H}^{(i)}_{11}$ at the receiver and a noiseless CSI feedback with a delay of a single channel use. Since $\mathbf{H}^{(i)}$ unitary, $\mathbf{H}^{(i)}_{21}$ can be computed from $\mathbf{H}^{(i)}_{11}$ and we assume that the receiver noiselessly communicates $\mathbf{H}^{(i)}_{21}$ to the transmitter%\footnote{Alternatively, $\mathbf{H}^{(i)}_{11}$ could have been communicated back to the transmitter, however, $\mathbf{H}^{(i)}_{21}$ bears less information.}
. Note that $\mathbf{H}^{(i)}_{21}$ completes $\mathbf{H}^{(i)}_{11}$ into orthonormal vectors, thus for $m_t+m_r-m>1$ and certain matrix instantiations, the computed $\mathbf{H}^{(i)}_{21}$ is not unique and can be chosen wisely (see {Remark \ref{Uniqueness}}).

Now, let the transmitter excites the following signal from the addressed modes at each channel use $i=1,\ldots,n$
\[\mathbf{\underline{x}}^{(i)}=\begin{bmatrix}
\mathbf{x}^{(i)}_1\\
\vdots\\
\mathbf{x}^{(i)}_{m_t+m_r-m}\\
\mathbf{H}^{(i-1)}_{21}\mathbf{\underline{x}}^{(i-1)}
\end{bmatrix}~.\]
That is, the transmitter conveys $m_t+m_r-m$ new information bearing symbols and $\mathbf{H}^{(i-1)}_{21}\mathbf{\underline{x}}^{(i-1)}$, a linear combination of the signal that was transmitted in the previous channel use ($\mathbf{\underline{x}}^{(0)}$ is a vector of zeros). Note that $\mathbf{H}$ is unitary, thus the power constraint is left satisfied.

We shall now assume that after the last signal $\mathbf{\underline{y}}^{(n)}$ is received, the receiver gets as a side information the following noisy measures
\begin{equation}
{\mathbf{\underline{y}}}_{\text{si}}=\sqrt{\rho}\mathbf{H}^{(n)}_{21}\mathbf{\underline{x}}^{(n)}+{\mathbf{\underline{z}}}_{\text{si}} ~, \label{scheme2}
\end{equation}
where the components of ${\mathbf{\underline{z}}}_{\text{si}}$ are i.i.d. $\mathcal{CN}(0,1)$. Thus the receiver can linearly combine $\underline{y}^{(n)}$ and ${\mathbf{\underline{y}}}_{\text{si}}$ in the following manner
\begin{equation}
\tilde{\mathbf{\underline{y}}}^{(n)}=\begin{bmatrix} \mathbf{H}^{(n)\dag}_{11} &  \mathbf{H}^{(n)\dag}_{21}\end{bmatrix}   \begin{bmatrix}[c]~\mathbf{\underline{y}}^{(n)}\\{\mathbf{\underline y}}_{\text{si}}\end{bmatrix}~ \label{scheme1}
\end{equation}
to yield
\begin{align}
\tilde{\mathbf{\underline{y}}}^{(n)}&= \sqrt{\rho}\mathbf{\underline{x}}^{(n)}+\tilde{\mathbf{\underline{z}}} \label{scheme3}
\end{align}
where the entries of $\tilde{\mathbf{\underline{z}}}$ are i.i.d. $\mathcal{CN}(0,1)$. We remind that the first $m_t+m_r-m$ entries of $\mathbf{\underline{x}}^{(n)}$ are new information bearing symbols and the last entries are equal to $\mathbf{H}^{(n-1)}_{21}\mathbf{\underline{x}}^{(n-1)}$. Thus, the last $m-m_r$ entries of $\tilde{\mathbf{\underline{y}}}^{(n)}$, denoted $\breve{\mathbf{\underline{y}}}$, satisfy
\[\breve{\mathbf{\underline{y}}}=\sqrt{\rho}\mathbf{H}^{(n-1)}_{21}\mathbf{\underline{x}}^{(n-1)}+\breve{\mathbf{\underline{z}}}~.\]
where $\breve{\mathbf{\underline{z}}}$ are the last $m-m_r$ entries of $\tilde{\mathbf{\underline{z}}}$. Again, the receiver can linearly combine $\underline{y}^{(n-1)}$ and $\breve{\mathbf{\underline{y}}}$ as
\begin{equation}
\tilde{\mathbf{\underline{y}}}^{(n-1)}=\begin{bmatrix} \mathbf{H}^{(n-1)\dag}_{11} &  \mathbf{H}^{(n-1)\dag}_{21}\end{bmatrix}   \begin{bmatrix}[c]~\mathbf{\underline{y}}^{(n-1)}\\\breve{\mathbf{\underline{y}}}\end{bmatrix}~
\end{equation}
to yield measures of $\mathbf{\underline{x}}^{(n-1)}$ as in Eq. \eqref{scheme3}. Repeating this procedure for $i=n-1~\rightarrow~1$ results in $m_t+m_r-m$ independent streams of measures
\[\begin{bmatrix}\tilde{\mathbf{y}}^{(1)}_1 \\ \vdots \\ \tilde{\mathbf{y}}^{(1)}_{m_t+m_r-m} \end{bmatrix} ,~\ldots,~ \begin{bmatrix}\tilde{\mathbf{y}}^{(n)}_1 \\ \vdots \\ \tilde{\mathbf{y}}^{(n)}_{m_t+m_r-m} \end{bmatrix}~.\]

The scheme above is feasible if the side information after channel use $n$ is being conveyed by the transmitter through a neglectable number of channel uses (with respect to $n$, see Remark \ref{ClosingSymbols}). In that case the receiver can construct $m_t+m_r-m$ independent SISO channels, each with signal-to-noise ratio $\rho$. Thus the scheme supports a rate arbitrarily close (as $n$ is larger) to $(m_t+m_r-m)\log(1+\rho)$ with zero outage probability. Note that the scheme essentially completes the singular values of the channel to 1. This is feasible since $m-m_r<m_t$, thus at each channel use the transmitter can transmit $\mathbf{H}^{(i-1)}_{21}\mathbf{\underline{x}}^{(i-1)}$, a signal of $m-m_r$ entries, and new symbols.

The scheme presented above can be easily expanded to the case where the feedback delay is $l$ channel uses. In that case the transmitter conveys at each channel use $m_t+m_r-m$ new information bearing symbols and $\mathbf{H}^{(i-l)}_{21}\mathbf{\underline{x}}^{(i-l)}$, a linear combination of the signal that was transmitted $l$ channel uses before. After channel use $n$, the transmitter would have to convey $l$ noisy measures of the last $l$ signals, so that the receiver could construct $m_t+m_r-m$ independent SISO channels. This can be done in a fixed number of channel uses (see Remark \ref{ClosingSymbols}), thus as $n$ is larger, the transmission rate of the scheme approaches $(m_t+m_r-m)\log(1+\rho)$.

\begin{rem} [Outdated feedback]
Our scheme exploits a noiseless CSI feedback system to communicate a (possibly) outdated information - the channel realization in previous channel uses. Thus, the feedback is not required to be fast, that is, no limitations on the delay time $l$. However, if $l$ is smaller than the coherence time of the channel, the feedback may carry information about the current channel realization. Thus, the transmitter can exploit the up-to-date feedback to use more efficient schemes. Nevertheless, for systems with a long delay time (e.g., relatively long distance optical fibers), the channel can be regarded as non-ergodic with an outdated feedback. In these cases our scheme efficiently achieves zero outage probability.  %Nevertheless, the delay has implications on the system latency, storage size etc.
\end{rem}

\begin{rem}[Simple decoding]
The scheme linearly process the received signals to construct $m_t+m_r-m$ independent streams of measures, each with signal-to-noise $\rho$. This allows the decoding stage to be simple, where a SISO channel decoder can be used, removing the need for further MIMO signal processing.
\end{rem}

\begin{rem} [Side information measures] \label{ClosingSymbols}
For a feedback with a delay of $l$ channel uses, the transmitter has to convey $\mathbf{H}^{(i)}_{21}\mathbf{\underline{x}}^{(i)}$, for each $i=n-(l-1),\ldots,n $, such that the receiver can extracted a vector of noisy measures with signal-to-noise ratio that is not smaller than $\rho$. This is feasible with a finite number of channel uses.
For example, the repetition scheme can be used to convey these measures (see Section \ref{sec_DMT} Example \ref{DMT_example_repetitionScheme}). Suppose each $\mathbf{H}^{(i)}_{21}\mathbf{\underline{x}}^{(i)}$ is conveyed to the receiver within $N_{si}$ channel uses (e.g., for the repetition scheme $N_{si}=m_t(m-m_r)$). By taking large enough $n$ (with respect to $l\cdot N_{si}$) one can approach the rate $(m_t+m_r-m)\log(1+\rho)$.
\end{rem}

\begin{rem} [Uniqueness of $\mathbf{H}_{21}$] \label{Uniqueness}
The scheme can be further improved to support even an higher data rate with zero outage probability. For example, the last $m-m_r$ entries of the transmitted signal at the first channel use can be used to excite information bearing symbols instead of the zeros symbols. Furthermore, as was mentioned above, when $m_t+m_r-m>1$, $\mathbf{H}^{(i)}_{21}$ is not unique; there are many $(m-m_r)\times m_t$ matrices that complete the columns of $\mathbf{H}^{(i)}_{11}$ into orthonormal vectors. Thus, the receiver can choose $\mathbf{H}^{(i)}_{21}$ to be the one with the largest number of zeros rows. Now, at time $i+1$ the transmitter excites $m_t+m_r-m$ new symbols and $\mathbf{H}^{(i)}_{21}\mathbf{\underline{x}}^{(i)}$, a retransmission of $\mathbf{\underline{x}}^{(i)}$, the transmitted signal at time $i$. With an appropriate choice of $\mathbf{H}^{(i)}_{21}$, $\mathbf{H}^{(i)}_{21}\mathbf{\underline{x}}^{(i)}$ contains entries that are zero. Instead, these entries can contain additional new information bearing symbols. An open question is how to further enhance the data rate. One would like to exploit the feedback to approach the empirical capacity for any realization of $\mathbf{H}_{11}$. Note that this rate is achievable with an up-to-date feedback. Further approaching this rate with an outdated feedback system (and with zero outage probability) is left for future research.
\end{rem}

\section{Diversity Multiplexing Tradeoff} \label{sec_DMT}
Using multiple modes/antennas is an important mean to improve performance in optical/wireless systems. The performance can be improved by increasing the transmission rate or by reducing the error probability. A coding scheme can achieve both performance gains, however there is a fundamental tradeoff between how much each can get. This tradeoff is known as the diversity-multiplexing tradeoff (DMT). The optimal tradeoff for the Rayleigh fading channel was found in \cite{DMT}. In this section we seek to find the optimal tradeoff for the Jacobi channel.

To better understand the concepts of diversity and multiplexing gains in the Jacobi channel we start with the following example.
\begin{example}[Repetition scheme] \label{DMT_example_repetitionScheme}
Suppose the transmitter excites the following ($m_t$ entries) signals in each $m_t$ consecutive channel uses:
%\[\begin{bmatrix} \mathbf{x}  & 0 & \ldots & 0 \\0 & \mathbf{x} &\ldots & 0 \\ & & \vdots & & \\ 0 & 0 &\ldots &  \mathbf{x}\end{bmatrix}~.\]
\[\begin{bmatrix} \mathbf{x} \\ 0 \\ \vdots \\ 0\end{bmatrix},\begin{bmatrix} 0\\ \mathbf{x} \\ \vdots \\ 0\end{bmatrix}, \ldots, \begin{bmatrix} 0\\ 0 \\ \vdots \\\mathbf{x}\end{bmatrix}~.\]
Let us make the simplifying assumptions that $\mathbf{x}$ is an uncoded QPSK symbol and that $m_t\leq m_r$ (similar results can be obtained also for $m_t> m_r$ and for higher constellation sizes). We further assume that the channel realization is known at the receiver and is constant within the $m_t$ channel uses. It can be shown that in that case the average error probability satisfies
\begin{align}
P_e\big(\rho\big)&\doteq \mathbb{E}[\exp\big({-\frac{\rho}{2}\sum_{i=1}^{m_t}\lambda_i}\big)]~, \label{errPrRep1}
\end{align}
where the expectation is over $\{\lambda_i\}_{i=1}^{m_t}$, the eigenvalues of $\mathbf{H}_{11}^\dag \mathbf{H}_{11}$. Here and throughout the rest of the paper we use $\doteq$ to denote \emph{exponential equality}, i.e., $f(\rho)\doteq \rho^d$ denotes
\begin{equation}
\lim_{\rho\rightarrow \infty}\frac{\log f(\rho)}{\log \rho}=d~.
\end{equation}
Now, for $m_t+m_r\leq m$, we can apply the joint distribution of the {\emph{unordered}} eigenvalues of a Jacobi matrix $\mathcal{J}(m_r,m-m_r,m_t)$, to write
{{\begin{align}
P_e\big(\rho\big)&\doteq \tfrac{K^{-1}_{m_t,m_r,m}}{m_t!}\int_0^1 \ldots \int_0^1 \prod_{i=1}^{m_t} \lambda_i^{m_r-m_t}(1-\lambda_i)^{m-m_r-m_t}e^{-\tfrac{\rho}{2}\lambda_i}\prod_{i<j}(\lambda_j-\lambda_i)^2 \prod_{i=1}^{m_t}d\lambda_i~. \label{errPrRep2}
\end{align}}}
Note that the term
\[\prod_{1\leq i <j\leq m_t}(\lambda_j-\lambda_i)\]
is the determinant of the Vandermonde matrix
\[\begin{bmatrix} 1 & \ldots & 1\\ \lambda_1 & \ldots & \lambda_{m_t}\\ \vdots & & \vdots\\ \lambda_1^{m_t-1} & \ldots & \lambda_{m_t}^{m_t-1}\end{bmatrix}~.\]
Thus we can write
\begin{align}
\prod_{1\leq i <j\leq m_t}(\lambda_j-\lambda_i)^2=\sum_{\sigma_1,\sigma_2\in S_{m_t}} (-1)^{sgn(\sigma_1)+sgn(\sigma_2)}\prod_{i=1}^{m_t}\lambda_i^{\sigma_1(i)+\sigma_2(i)-2}~, \label{errPrRep3}
\end{align}
where $S_{m_t}$ is the set of all permutations of $\{1,\ldots,m_t\}$ and $sgn(\sigma)$ denotes the signature of the permutation $\sigma$. Applying \eqref{errPrRep3} into \eqref{errPrRep2} results
\begin{align}
P_e\big(\rho\big)\doteq \tfrac{K^{-1}_{m_t,m_r,m}}{m_t!}\sum_{\sigma_1,\sigma_2\in S_{m_t}} (-1)^{sgn(\sigma_1)+sgn(\sigma_2)} \prod_{i=1}^{m_t}\int_0^1  &\lambda_i^{m_r-m_t+\sigma_1(i)+\sigma_2(i)-2}\times \nonumber\\
&\times (1-\lambda_i)^{m-(m_r+m_t)}e^{-\tfrac{\rho}{2}\lambda_i} d\lambda_i~.
\end{align}
It can be further shown that the right-hand-side of above is dominated (for large $\rho$) by the following term
\begin{align}\tfrac{K^{-1}_{m_t,m_r,m}}{m_t!}\sum_{\sigma_1,\sigma_2\in S_{m_t}} (-1)^{sgn(\sigma_1)+sgn(\sigma_2)}\prod_{i=1}^{m_t} (m_r-m_t+\sigma_1(i)+\sigma_2(i)-2)!(\tfrac{\rho}{2})^{-(m_r-m_t+\sigma_1(i)+\sigma_2(i)-1)}~.\end{align}
Thus, for $m_t+m_r\leq m$, the average error probability satisfies
\begin{align}
P_e\big(\rho\big)&\doteq \rho^{-\sum_{i=1}^{m_t}(m_r-m_t+2i-1)}\\
&\doteq\rho^{-m_r  m_t}~. \label{repScheme_firstCase}
\end{align}
For $m_t+m_r>m$, by applying Lemma \ref{lem1} into \eqref{errPrRep1} we get
\[P_e\big(\rho\big)\doteq e^{-\tfrac{\rho(m_t+m_r-m)}{2}} \mathbb{E}[\exp\big({-\tfrac{\rho}{2}\sum_{i=1}^{m-m_r}\tilde{\lambda}_i}\big)]~,\]
where $\{\tilde{\lambda}_i\}_{i=1}^{m-m_r}$ are the eigenvalues of $\mathbf{H}_{22} \mathbf{H}_{22}^\dag$. Thus, we can conclude that the error probability of the repetition scheme satisfies
\begin{equation}  \label{errPrRepFin}
P_e\big(\rho\big)\doteq
\left\{
\begin{array}{ll}
\rho^{-m_r  m_t} &,~ m_t+m_r\leq m\\
e^{-\tfrac{\rho(m_t+m_r-m)}{2}}  \rho^{-(m-m_t)  (m-m_r)} &,~ m_t+m_r> m ~.
\end{array} \right.
\end{equation}
\end{example}

\begin{figure}[ht]
\center
\epsfig{file=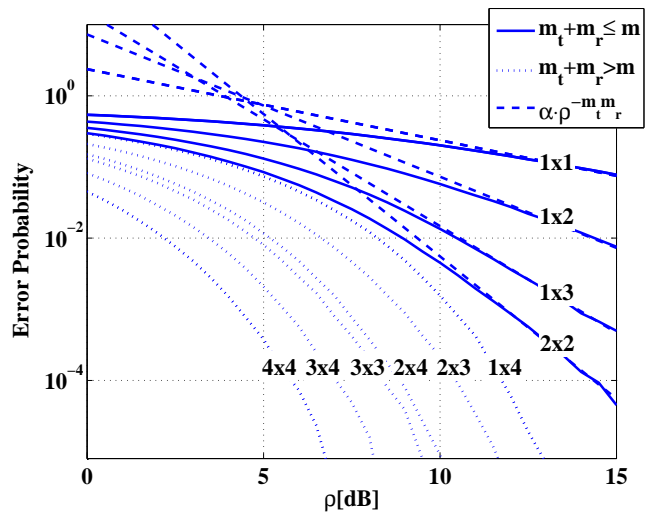,height=0.43\textwidth,width=0.55\textwidth,clip=}
\caption{The average error probability of the repetition scheme vs. $\rho$, for fixed number of supported modes $m=4$ and various numbers of transmit$\times$receive modes. The error probability curves are depict in solid lines for systems satisfying $m_t+m_r\leq m$ and in dotted lines for systems satisfying $m_t+m_r>m$. The dashed lines are given to emphasize the decaying order of the non-exponentially decaying curves.} \label{errorPr_n4}
\end{figure}

In Fig. \ref{errorPr_n4} we present the average error probability vs. $\rho$ for $m=4$ and various combinations of $m_t,~m_r$ (the error probability is symmetric in $m_t,~m_r$, thus all combinations of $m_t,~m_r$ are plotted). Note the decaying order of the curves and how they turn exponentially decaying when $m_t+m_r>m$.

Eq. \eqref{errPrRepFin} implies that when using $m_t$ transmit and $m_r$ receive modes, where $m_t+m_r\leq m$, the exponent of the dominant term in the average error probability is $-m_r  m_t$. Comparing to a system with a single transmit and a single receive mode, the decaying order of the average error probability is improved by a factor of $m_r m_t$. This gain is termed {\emph{diversity gain}}. When enough modes are being addressed by the transmitter and the receiver to satisfy $m_t+m_r> m$, we get an average error probability that exponentially decays with $\rho$; that is, an unbounded diversity gain. Thus, as more modes are being addressed, the diversity gain of the repetition scheme is greater. Since the total transmitted power is spread over all $m$ available modes, addressing only some modes at the receiver results in a power loss. As the number of these modes is larger, the probability for a substantial power loss is smaller; hence, smaller error probability. As the signal is transmitted from more modes, the average power in each receive mode is larger since the propagation paths are orthogonal. This is in analogy to the Rayleigh channel where as the signal passes through more (independent) faded paths, the decaying order of the error probability increases. However, it turns out that in the Jacobi channel there is a transition threshold in which enough modes are being addressed to ensure a certain received power. This results in an exponentially decaying error probability for certain rates.

Now, using multiple modes can also improve the data rate of the system. In the example above the rate is fixed, $R(\rho)=1/m_t ~(\text{bps/Hz})$ for any $\rho$. Increasing the data rate with $\rho$ to support a rate of $R(\rho)=r\log\rho ~(\text{bps/Hz})$ for some $0<r<1/m_t$, can be achieved by increasing the constellation size of the transmitted signal. In that case the data rate is improved by a factor of $r$ comparing to a system with a single transmit and a single receive mode. This gain is termed {\emph{multiplexing gain}} \footnote{The multiplexing gain in the given example is 0.}. By increasing the constellation size, however, the minimum distance between the constellation points decreases, resulting an error probability with a smaller decaying order; that is, a smaller diversity gain. Thus, there is a tradeoff between diversity and multiplexing gains.

We now turn to analyze the DMT in the Jacobi model. To that end, we formalize the concepts of {{diversity gain}} and {{multiplexing gain}} by quoting some definitions from \cite{DMT} \footnote{Note that in \cite{DMT} the definitions in \ref{def_DMT} were made with respect to the average signal-to-noise ratio at each receive mode, denoted $\bar{\rho}$. However, since $\bar{\rho}=\rho\mathbb{E}[{\emph{~tr}}(\mathbf{H}_{11}Q\mathbf{H}_{11}^\dag)]/m_r$, where $Q$ is the transmitted covariance matrix, we can write
%\begin{equation*}
$\lim_{\bar{\rho}\rightarrow \infty}\log \bar{\rho}=\lim_{\rho\rightarrow \infty}\log \rho~.$
%\end{equation*}
Hence the definitions in \ref{def_DMT} coincide with those in \cite{DMT}.}.
\begin{defn} \label{def_DMT}
Let a scheme be a family of codes $\big\{\mathcal{C}(\rho)\big\}$ of block length $l$, one at each $\rho$ level. Let $R(\rho)$ (bps/Hz) be the rate of the code $\mathcal{C}(\rho)$. A scheme $\big\{\mathcal{C}(\rho)\big\}$ is said to achieve spatial multiplexing gain $r$ and diversity gain $d$ if the data rate satisfies
\[\lim_{\rho\rightarrow \infty} \frac{R(\rho)}{\log \rho}=r\]
and the average error probability satisfies
\[\lim_{\rho\rightarrow \infty} \frac{P_e(\rho)}{\log \rho}=-d ~.\]
For each $r$, define $d^*(r)$ to be the supremum of the diversity advantage achieved over all schemes.
\end{defn}

\subsection{Case \Rmnum{1} - $m_t+m_r\leq m$}
The following Theorem provides the optimal DMT of a Jacobi channel with $m_t,~m_r$ and $m$ satisfying $m_t+m_r\leq m$. In \cite{DMT} it was shown that the average error probability in the high SNR regime (large $\rho$) is dominated by the outage probability. Furthermore, the outage probability for a transmission rate $R=r\log(1+\rho)$, where $r$ is integer, is dominated by the probability that $r$ singular values of the channel are $1$ and the other approach zero. We show that the distribution of the singular values of the Jacobi and Rayleigh channels are approximately identical near 0; essentially proving that the optimal tradeoff is identical in both models.

\begin{thm} \label{thm_DMT_firstCase}
Suppose $l\geq m_t+m_r-1$. The optimal DMT curve $d^*(r)$ for the channel defined in \eqref{channel_model_eq}, with $m_t,~m_r$ satisfying $m_t+m_r\leq m$, is given by the piecewise linear function that connects the points $(k,d^*(k))$ for $k=0,1,\cdots,m_{\text{min}}$, where
\begin{equation} \label{optimalCurve}
d^*(k)=(m_t-k)(m_r-k)~.
\end{equation}
\end{thm}
\begin{proof}
See Appendix \ref{Proof of Theorem_thm_DMT_firstCase}.\\
\end{proof}

\begin{figure}[t]
\center
\begin{subfigure}[t]{0.4\textwidth}
\includegraphics[height=0.78\textwidth,width=1\textwidth]{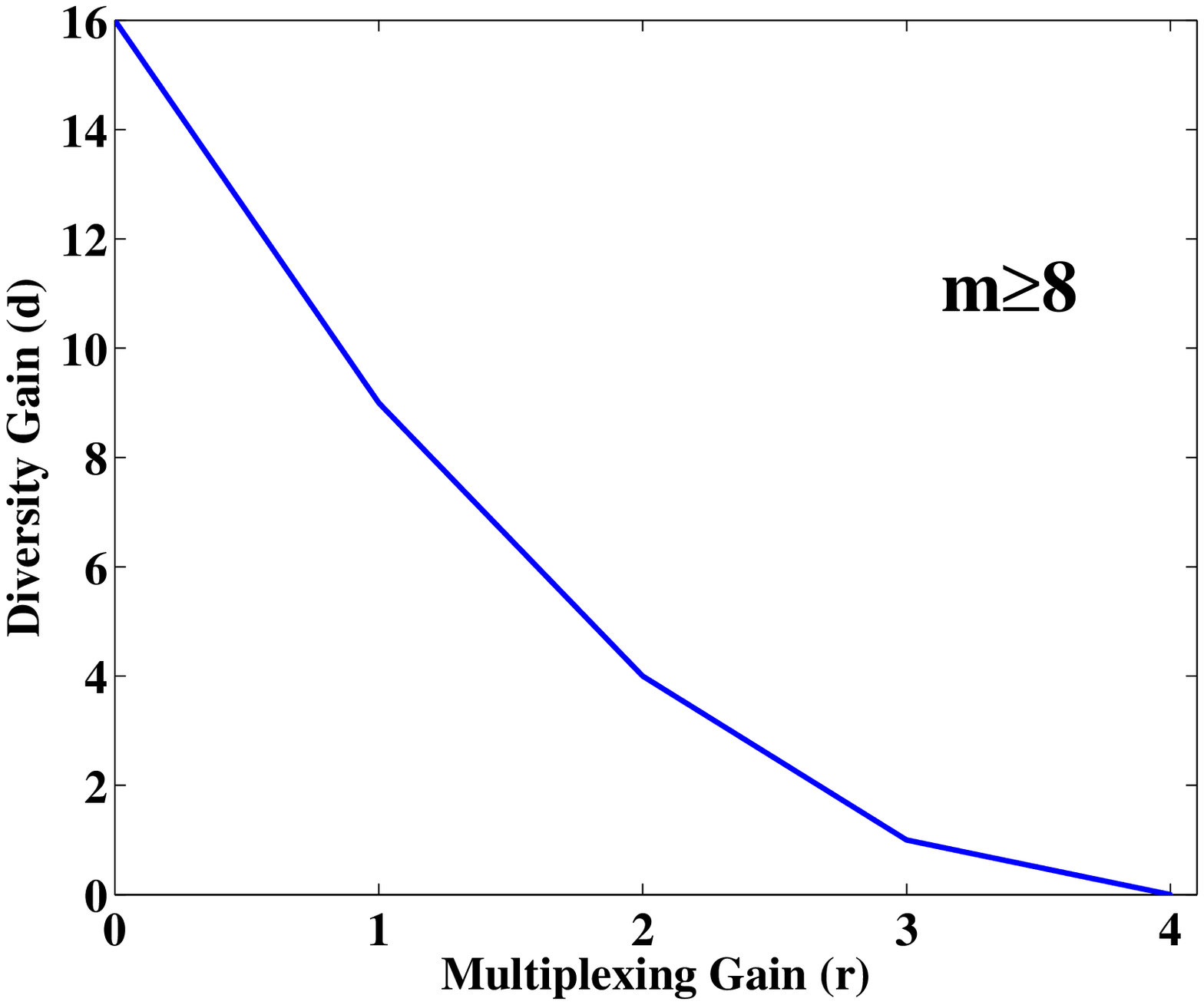}
\caption{} \label{DMT_n4_1}
\end{subfigure}\qquad
\begin{subfigure}[t]{0.4\textwidth}
\includegraphics[height=0.78\textwidth,width=1\textwidth]{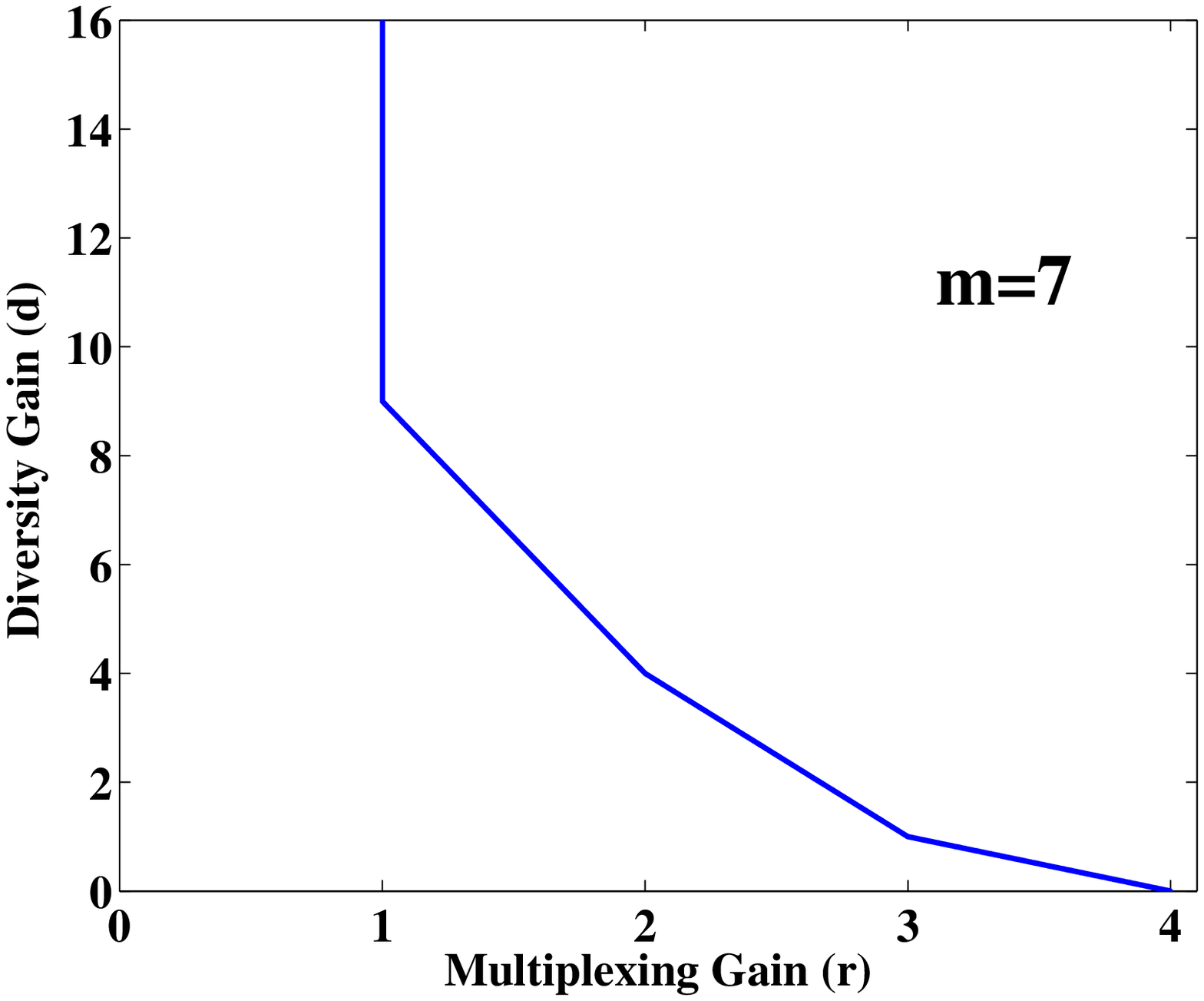}
\caption{} \label{DMT_n4_2}
\end{subfigure}\\
\center
\begin{subfigure}[t]{0.4\textwidth}
\includegraphics[height=0.78\textwidth,width=1\textwidth]{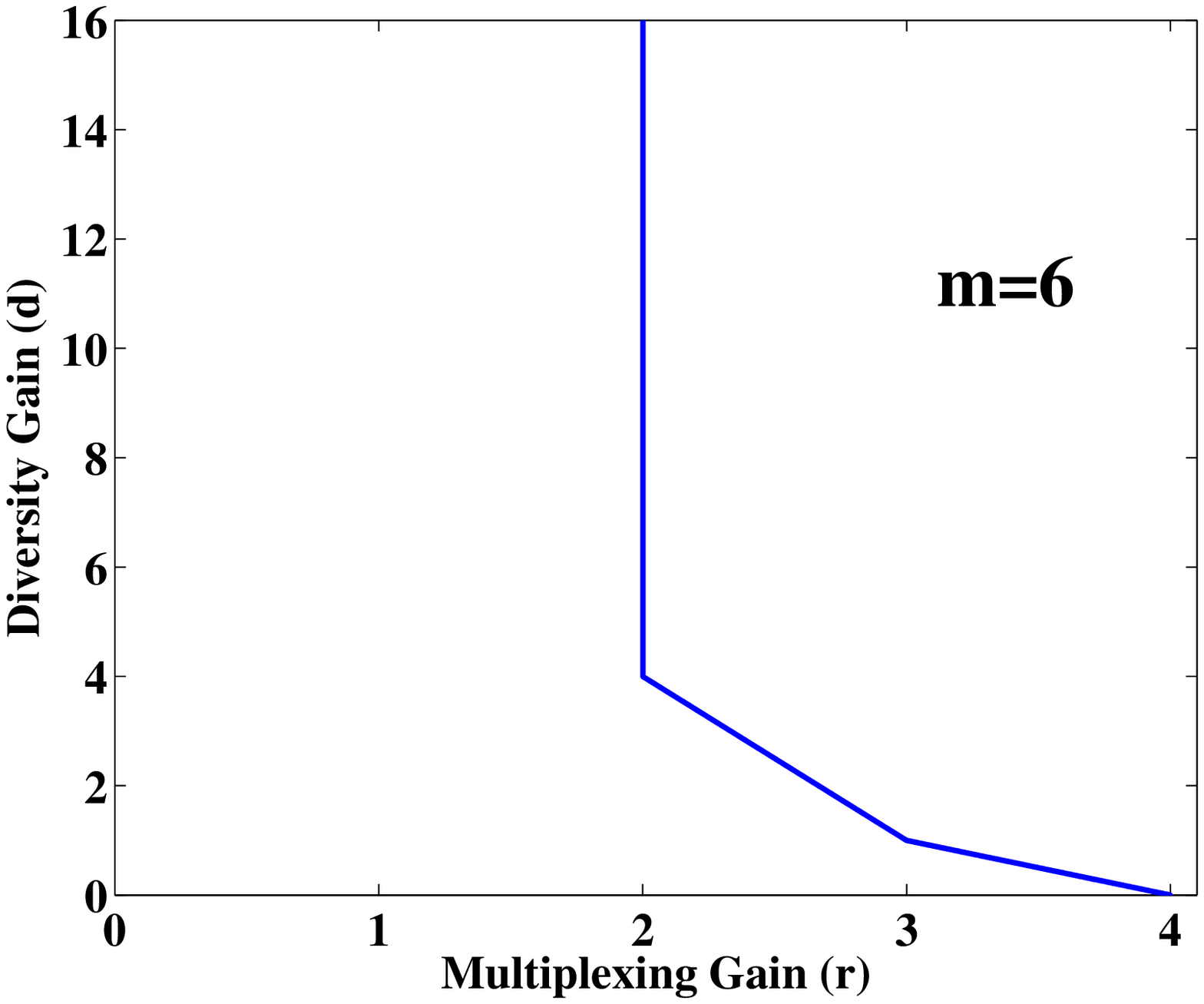}
\caption{} \label{DMT_n4_3}
\end{subfigure}\qquad
\begin{subfigure}[t]{0.4\textwidth}
\includegraphics[height=0.78\textwidth,width=1\textwidth]{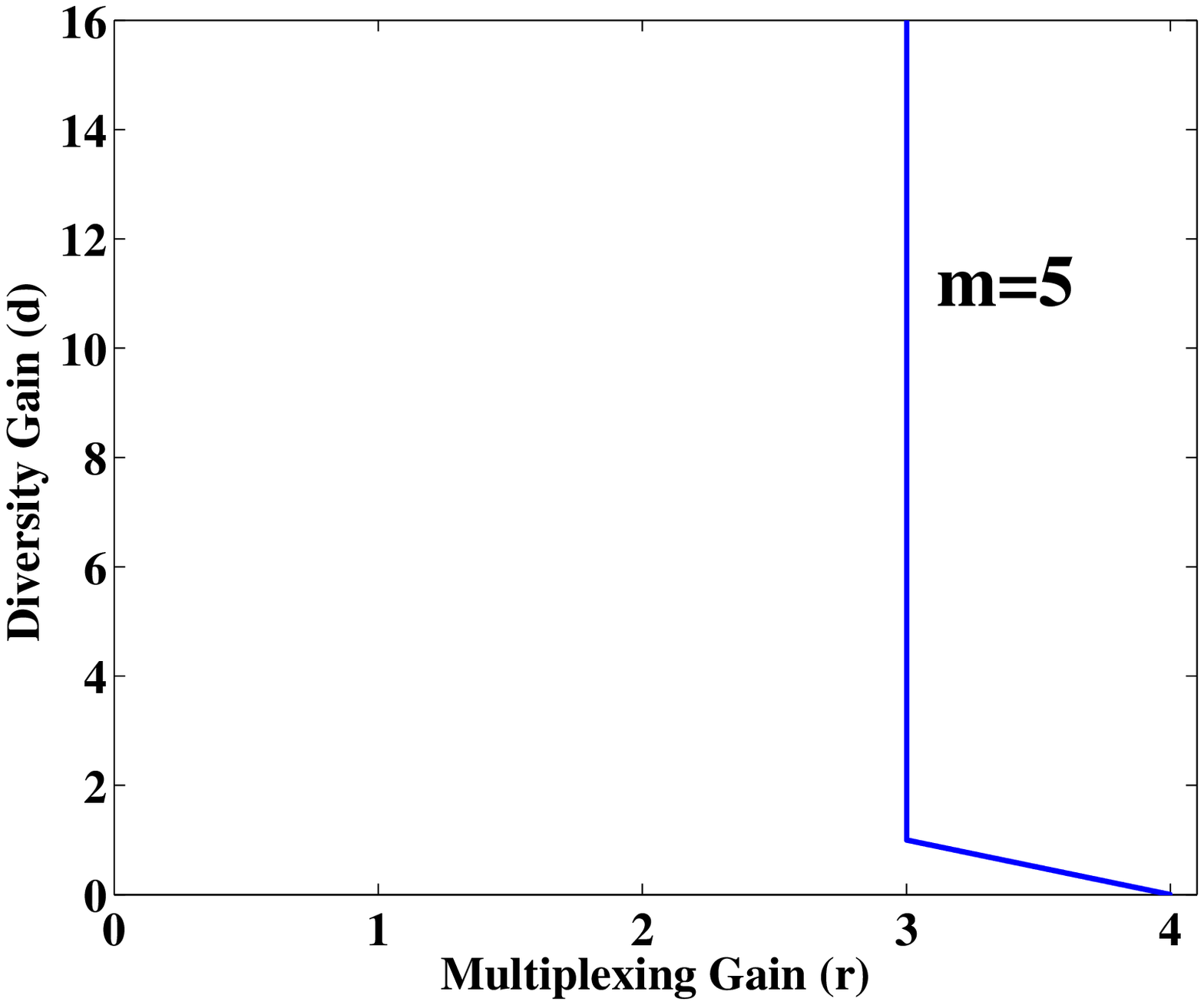}
\caption{} \label{DMT_n4_4}
\end{subfigure}\\
\center
\begin{subfigure}[t]{0.4\textwidth}
\includegraphics[height=0.78\textwidth,width=1\textwidth]{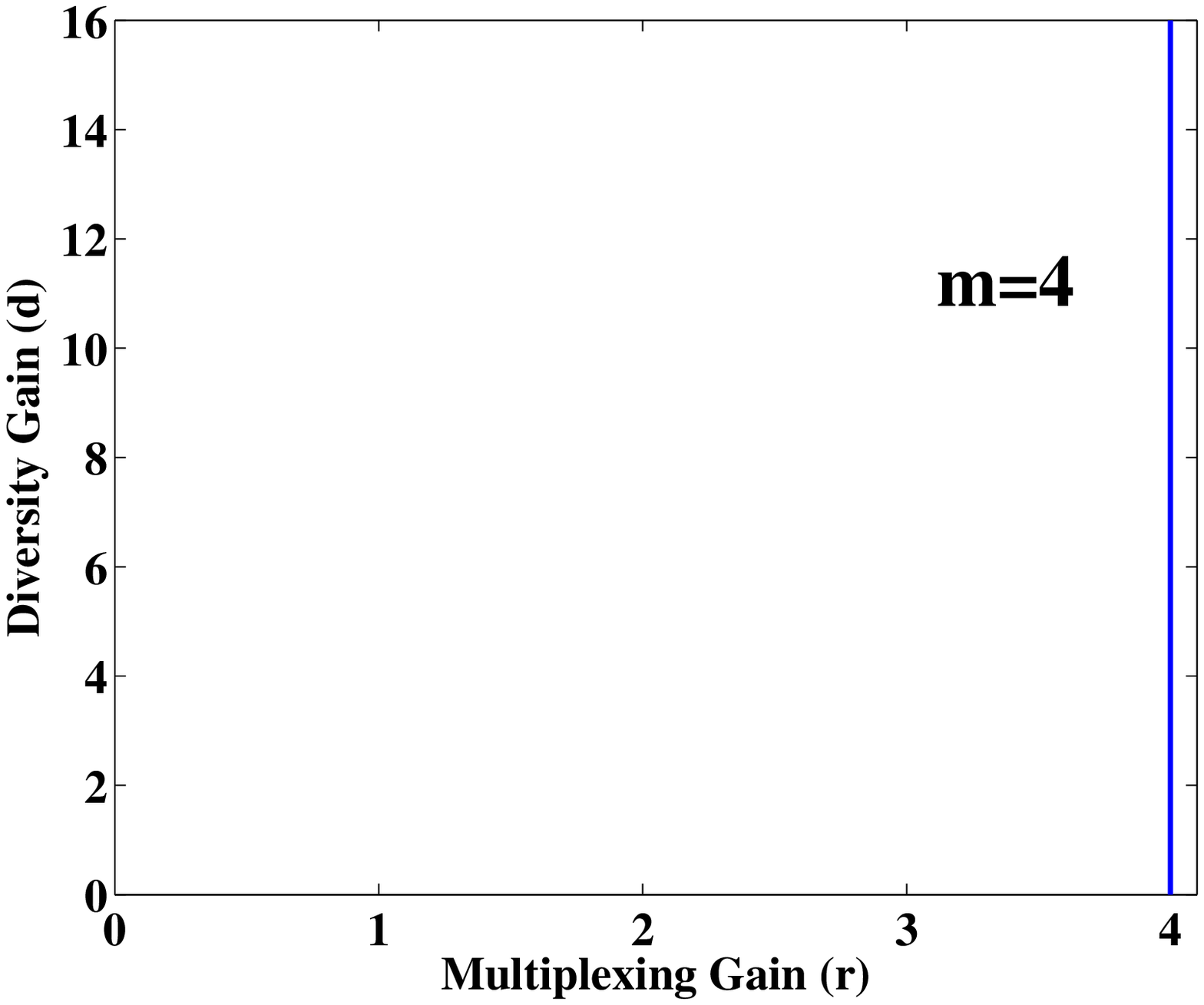}
\caption{} \label{DMT_n4_5}
\end{subfigure}
\caption{Optimal DMT curves for $m_t=m_r=4$, $l\geq7$ and various numbers of supported modes $m$.} \label{DMT_n4}
\end{figure}

Theorem \ref{thm_DMT_firstCase} suggests that                                                                                                                                                                                                              for $m_t+m_r\leq m$, the optimal DMT curve does not depend on $m$. Note that $m$ relates to the extent in which the elements of $\mathbf{H}_{11}$ are mutually independent -- the dependency is smaller as $m$ is larger. Hence, at high SNR (large $\rho$) the dependency between the path gains has no effect on the decaying order of the average error probability. Furthermore, the optimal DMT is identical to the optimal tradeoff in the analogous Rayleigh channel (where the path gains are independent).

\subsection{Case \Rmnum{2} - $m_t+m_r> m$}
According to Theorem \ref{thm_outagePr2case} a zero outage probability is achievable for rates below $(m_t+m_r-m)\log(1+\rho)$. Hence, for any $(m_t+m_r-m)>\delta>0$ there is a scheme $\big\{\mathcal{C}(\rho)\big\}$ with code rates $(m_t+m_r-m-\delta)\log(1+\rho)$ that achieves a zero outage probability; therefore, assuming $l$ is very large, achieves an exponentially decaying error probability. In that case the discussion about diversity is no longer of relevance. Nonetheless, one can think of the gain as infinite. This reveals an interesting difference between the Jacobi and Rayleigh channels - the maximum diversity gain is ``unbounded'' as opposed to $m_r  m_t$ in the later case.

\begin{thm} \label{thm_DMT_secondCase}
The optimal diversity multiplexing tradeoff curve $d^*(r)$ for the channel defined in \eqref{channel_model_eq}, with $m_t,~m_r$ satisfying $m_t+m_r> m$, is given by
{\small{\begin{equation} \label{DMT_curves_thm}
d^*(r)=\left\{
\begin{array}{ll}
d_{risdual}^*(r-(m_t+m_r-m)) &,~ r\geq m_t+m_r-m\\
\infty &,~ r< m_t+m_r-m~.
\end{array} \right.
\end{equation}}}
$d_{risdual}^*(r)$ is the optimal curve for a Jacobi channel with $m-m_r$ transmit and $m-m_t$ receive modes.
\end{thm}
\begin{proof}
At high SNR, in terms of minimal outage probability, we can take the covariance matrix of the transmitted signal to be $Q=\text{I}_{m_t}$, see Appendix \ref{Proof of Theorem_thm_DMT_firstCase}. Thus Theorem \ref{thm_outagePr2case} can be applied: for $r< m_t+m_r-m$ the minimal outage probability is zero hence the error probability turns exponentially decaying with $\rho$ (assuming $l$ is very large); for $r\geq m_t+m_r-m$ the outage probability equals the outage probability for $\tilde{r}=r-(m_t+m_r-m)$ in a system with $m-m_r$ transmit and $m-m_t$ receive modes. Noting that at high SNR the error probability is dominated by the outage probability (see Appendix \ref{Proof of Theorem_thm_DMT_firstCase}) completes the proof.\\
\end{proof}

Note that $d_{risdual}^*(r)$ in Eq. \eqref{DMT_curves_thm} is given by Theorem \ref{thm_DMT_firstCase} for any block length $l$ satisfying $l\geq m_t+m_r-1$. Fig. \ref{DMT_n4} depict the optimal DMT curve for $m_t=m_r=4$ and various numbers of supported modes $m$.

In the following example we try to illuminate the concept of infinite diversity gain.

\begin{figure}[t]
\center
\begin{subfigure}[t]{0.4\textwidth}
\includegraphics[height=0.78\textwidth,width=1\textwidth]{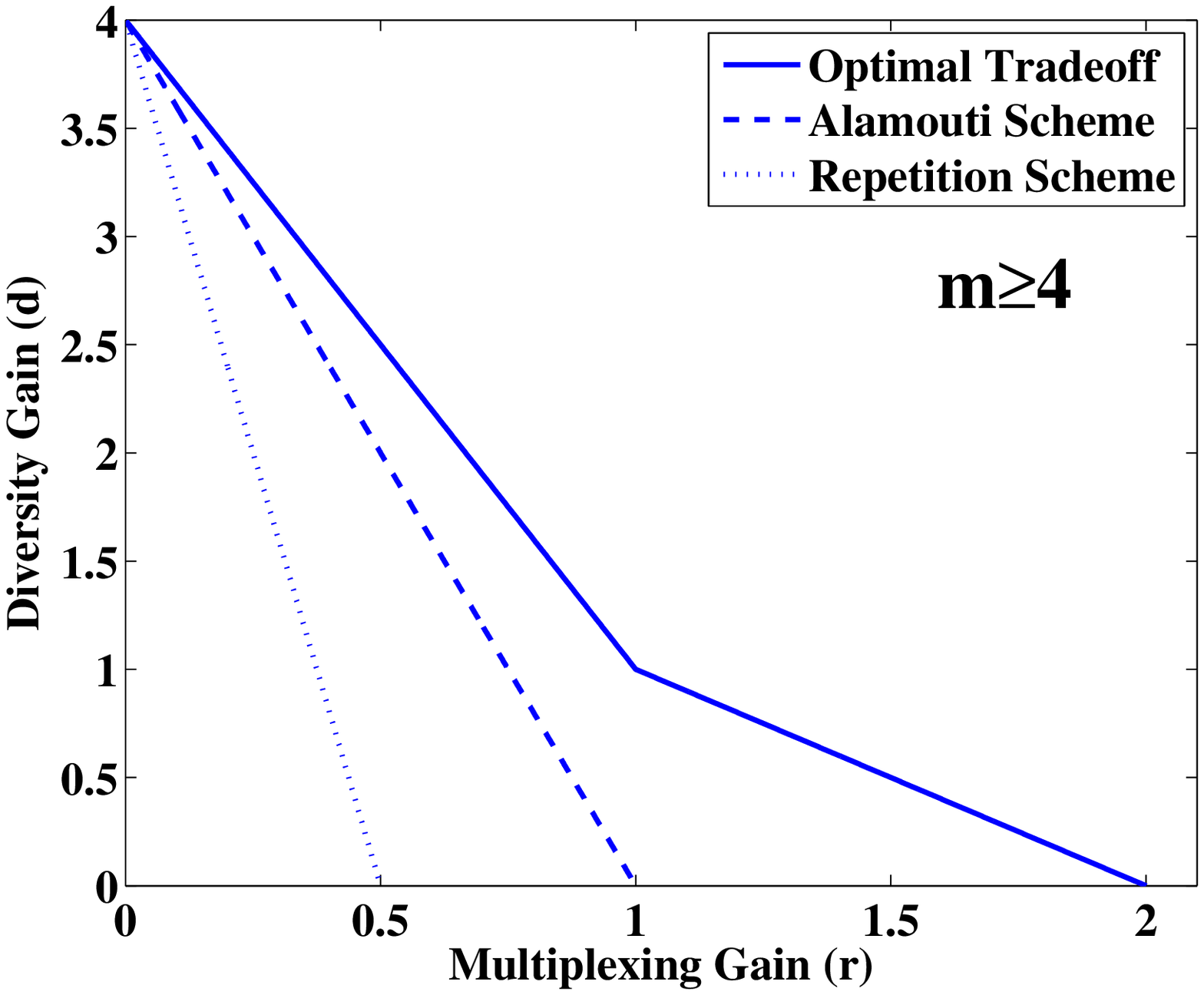}
\caption{} \label{DMT_n2_1}
\end{subfigure}\qquad
\begin{subfigure}[t]{0.4\textwidth}
\includegraphics[height=0.78\textwidth,width=1\textwidth]{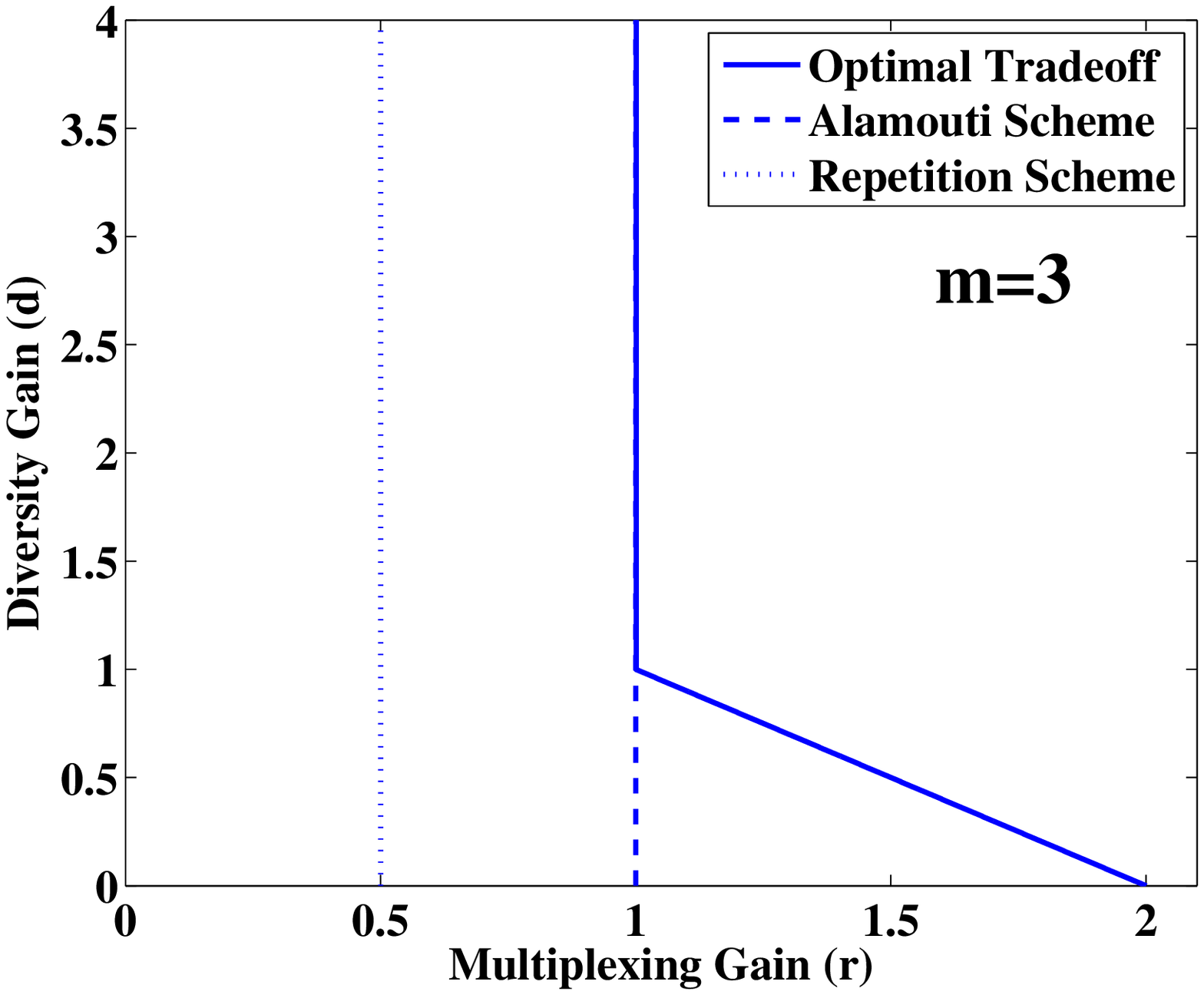}
\caption{} \label{DMT_n2_2}
\end{subfigure}\\
\center
\begin{subfigure}[t]{0.4\textwidth}
\includegraphics[height=0.78\textwidth,width=1\textwidth]{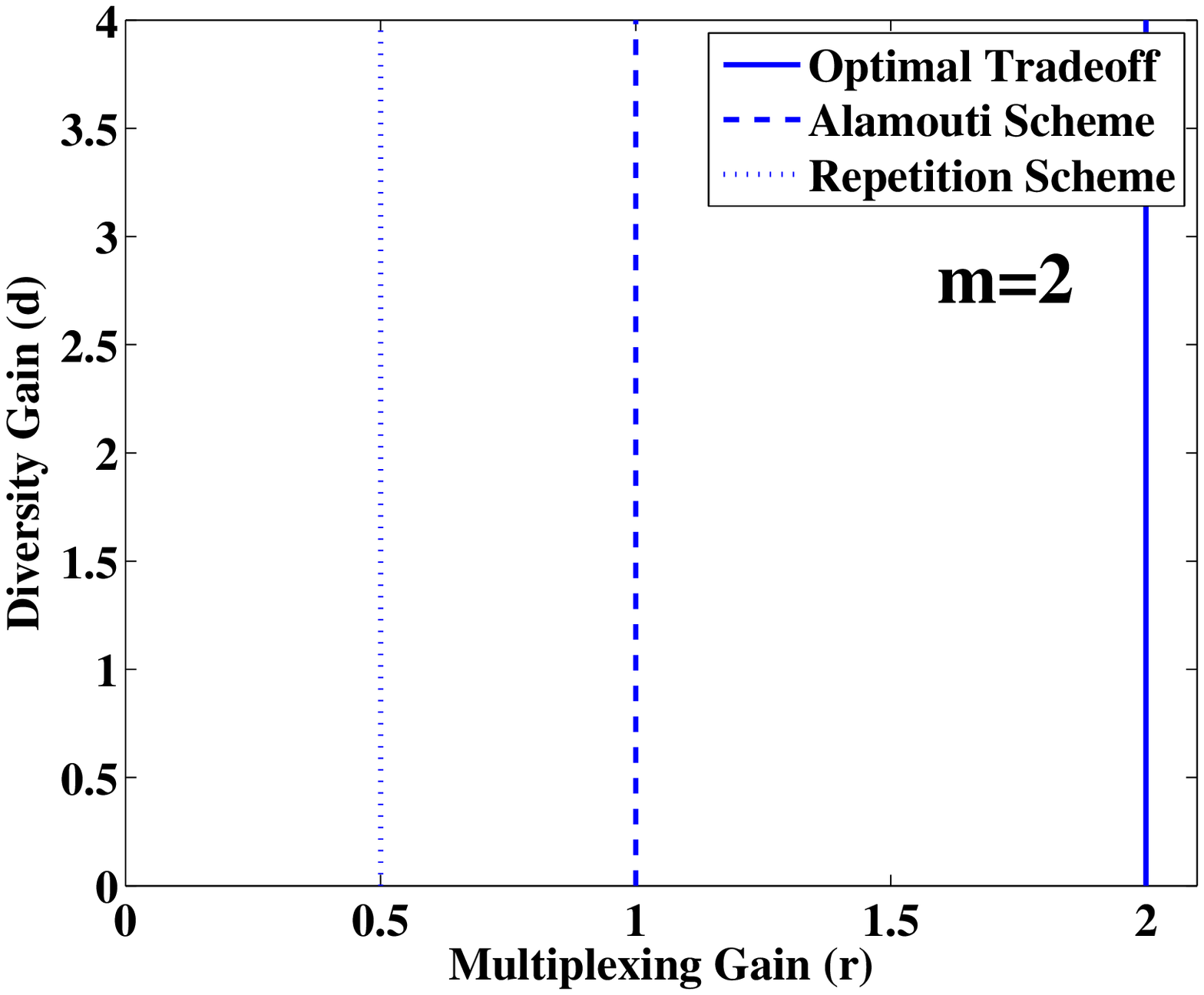}
\caption{} \label{DMT_n2_3}
\end{subfigure}
\caption{Comparison between Alamouti and the repetition scheme: $l\geq3$, $m_t=m_r=2$ and various numbers of supported modes $m$.} \label{DMT_n2}
\end{figure}

\begin{example} [$m_t=m_r=2$] \label{DMT_example_2times2}
We consider the $2 \times 2$ Alamouti scheme \cite{Alamouti}. Assuming a code block of length $l\geq3$ and rate $R=r\log\rho$ (bps/Hz), the transmitter excites in each two consecutive channel uses two information bearing symbols in the following manner:
%\[ \begin{bmatrix}[r] \mathbf{x}_1 & -\mathbf{x}_2^\dag \\ \mathbf{x}_2 &  \mathbf{x}_1^\dag\end{bmatrix}~.\]
\[ \begin{bmatrix}[r] \mathbf{x}_1 \\ \mathbf{x}_2 \end{bmatrix},~\begin{bmatrix}[r] -\mathbf{x}_2^\dag \\ \mathbf{x}_1^\dag\end{bmatrix}~.\]
ML decoding linearly combines the received measures and yields the following equivalent scalar channels:
\begin{equation}
\mathbf{y}_i=\sqrt{\|\mathbf{H}_{11}\|_F^2\rho}  \mathbf{x}_i+\mathbf{z}_i~,\quad \forall ~i=1,2
\end{equation}
where each $\mathbf{z}_i$ is i.i.d. $\mathcal{CN}(0,1)$ independent of $\mathbf{x}_i$ and $\mathbf{H}_{11}$. The probability for an outage event is given by
\begin{align}
P_{out}\text{\small{$(2,2,m;R)$}}&=Pr\big(\log(1+\|\mathbf{H}_{11}\|_F^2\rho)<r\log\rho\big)\\
&\doteq Pr\big(\|\mathbf{H}_{11}\|_F^2<\rho^{-(1-r)^+}\big)~.
\end{align}
Now, in the Rayleigh channel $\|\mathbf{H}_{11}\|_F^2$ is chi-square distributed with $2m_t m_r$ degrees of freedom. In thats case, as was shown in \cite{DMT}, the $2\times 2$ Alamouti scheme can achieve maximum diversity gain of $4$. However, in the Jacobi channel:
\begin{itemize}
\item for $m=2$ we have $\|\mathbf{H}_{11}\|_F^2=2$ ($\mathbf{H}_{11}=\mathbf{H}$ unitary).
\item for $m=3$ we have $\|\mathbf{H}_{11}\|_F^2\geq 1$ (by Lemma \ref{lem1}).
\item for $m\geq 4$ there is always a non-zero probability for an outage event.
\end{itemize}

Therefore, for $m=2$ and $m=3$, for any $r\leq 1$, we get equivalent unfading scalar channels with strictly zero outage probability and one can think of the maximum diversity gain as infinite. For $m\geq4$ it can be shown that the maximum diversity gain is $4$ and the DMT curve linearly connects the points $(1,0)$ and $(0,4)$.

In Example \ref{DMT_example_repetitionScheme} we saw that for multiplexing gain $r=0$ the repetition scheme achieves a diversity gain of $m_r m_t$ for systems satisfying $m_t+m_r\leq m$ and an unbounded gain for systems satisfying $m_t+m_r> m$. Thus, for $m\geq4$ the maximum diversity gain of this scheme is $4$ and it can be shown that the DMT curve linearly connects the points $(1/2,0)$ and $(0,4)$. For $m=2$ and $m=3$ we get an unbounded diversity gain for any multiplexing gain below $r=1/2$.

In Fig. \ref{DMT_n2} we compare these DMT curves to the optimal curves. Note that for $m=3$ the Alamouti scheme achieves the optimal DMT for $r=1$.
\end{example}

\section{Relation To The Rayleigh Model} \label{section_Rayleigh}

The Jacobi fading model is defined by the transfer matrix $\mathbf{H}_{11}$, a truncated $m_r\times m_t$ version of a Haar distributed $m\times m$ unitary matrix. We shall now examine the case where $m$ is very large with respect to $m_t$ and $m_r$.

Assuming $m_t\leq m_r$ and $m_t+m_r\leq m$, the statistics of the squared singular values of the Jacobi channel model follow the law of the Jacobi ensemble $\mathcal{J}(m_r,m-m_r,m_t)$. This ensemble can be constructed as
\begin{equation} \label{G_1_G_2}
\mathbf{G}_1^\dag \mathbf{G}_1(\mathbf{G}_1^\dag \mathbf{G}_1+\mathbf{G}_2^\dag \mathbf{G}_2)^{-1}~,
\end{equation}
where $\mathbf{G}_1$ and $\mathbf{G}_2$ are $m_r\times m_t$ and $(m-m_r)\times m_t$ independent Gaussian matrices. Thus, the squared singular values of $\mathbf{H}_{11}$ share the same distribution with the eigenvalues of \eqref{G_1_G_2}. Intuitively, in terms of the singularity statistics, the Jacobi channel can be viewed as an $m_r\times m_t$ sub-channel of an $m\times m_t$ {\emph{normalized}} Gaussian channel. Furthermore, for $m \gg m_r$ we have
\begin{align}
\mathbf{G}_1^\dag \mathbf{G}_1(\mathbf{G}_1^\dag \mathbf{G}_1+\mathbf{G}_2^\dag \mathbf{G}_2)^{-1} &= \mathbf{G}_1^\dag \mathbf{G}_1(\begin{bmatrix} \mathbf{G}_1\\ \mathbf{G}_2\end{bmatrix}^\dag \begin{bmatrix} \mathbf{G}_1\\ \mathbf{G}_2\end{bmatrix})^{-1} \label{sec_eq_Rayleigh}\\
&\approx \mathbf{G}_1^\dag \mathbf{G}_1(m \mathbb{E}[\underline{\mathbf{g}}  \underline{\mathbf{g}}^\dag])^{-1}\\
&= \tfrac{1}{m}\mathbf{G}_1^\dag \mathbf{G}_1~, \label{approx_ray_1}
\end{align}
where in \eqref{sec_eq_Rayleigh} we applied the law of large numbers ($\underline{\mathbf{g}}$ is a vector of $m_t$ independent components each distributed $\mathcal{CN}(0,1)$). In the same manner, for $m_t> m_r$, $m_t+m_r\leq m$ and $m \gg m_t$ the squared singular values of the Jacobi channel share the same distribution with the following ensemble of random matrices
\begin{equation} \label{approx_ray_2}
\mathbf{G}_1 \mathbf{G}_1^\dag(\mathbf{G}_1 \mathbf{G}_1^\dag+\mathbf{G}_2 \mathbf{G}_2^\dag)^{-1} \approx \tfrac{1}{m}\mathbf{G}_1 \mathbf{G}_1^\dag~.
\end{equation}
This allows us to conclude that up to a normalizing factor the Jacobi model approaches (with $m$) the Rayleigh model.

\begin{figure}[t]
\hspace{-0.8cm}
\begin{subfigure}[t]{0.5\textwidth}
\psfrag{aaa[dB]}{\textbf{\footnotesize{$\bar{{\rho}}$[dB]}}}
\epsfig{file=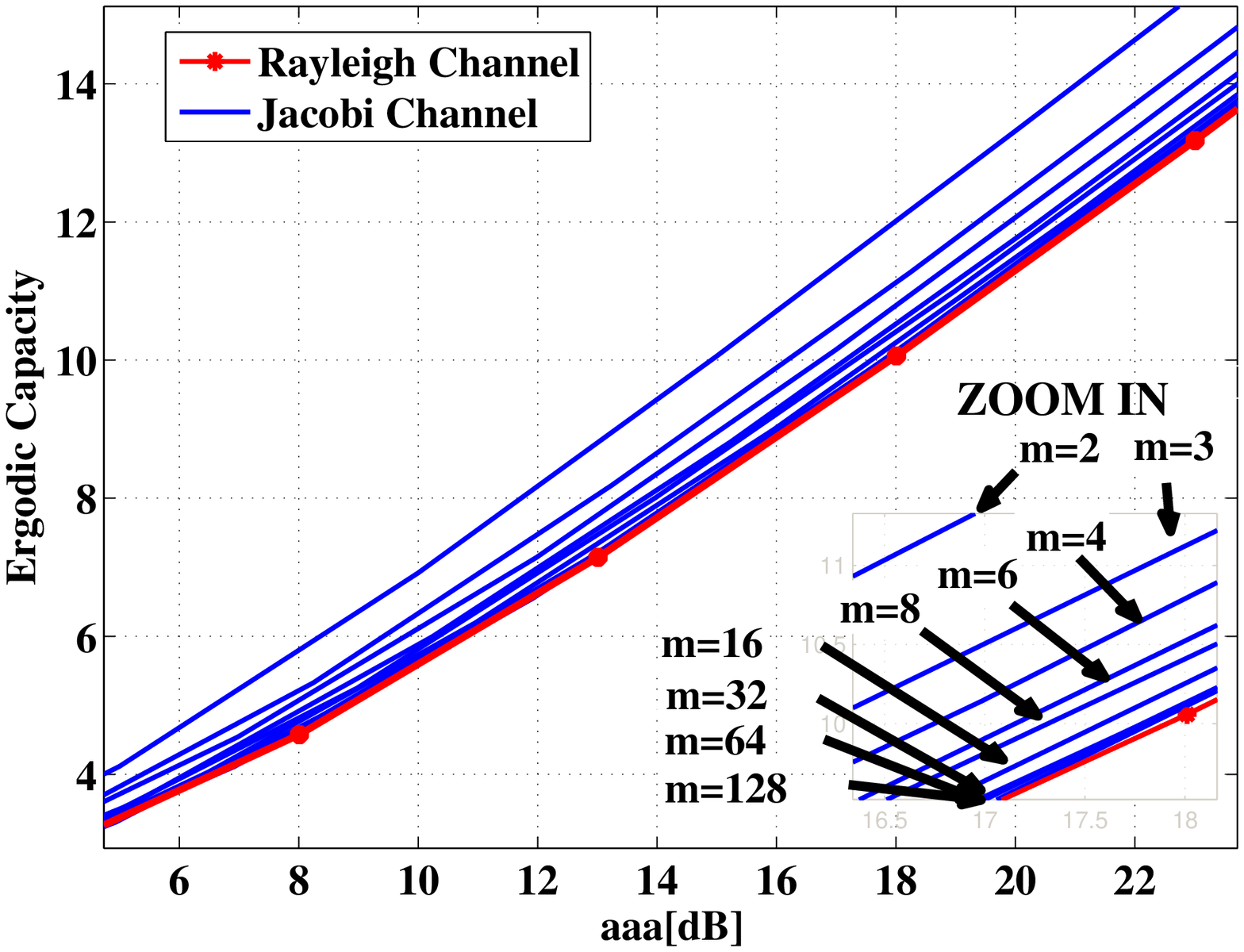,height=0.88\textwidth,width=1.1\textwidth,clip=}
\caption{} \label{Rayleigh_Jacobi_Figure_1}
\end{subfigure}\hspace{0.5cm}
\begin{subfigure}[t]{0.5\textwidth}
\psfrag{SNR}{\hspace{-0.5cm}{\textbf{\footnotesize{$r=R/\log(1+\bar{\rho})$}}}}
\epsfig{file=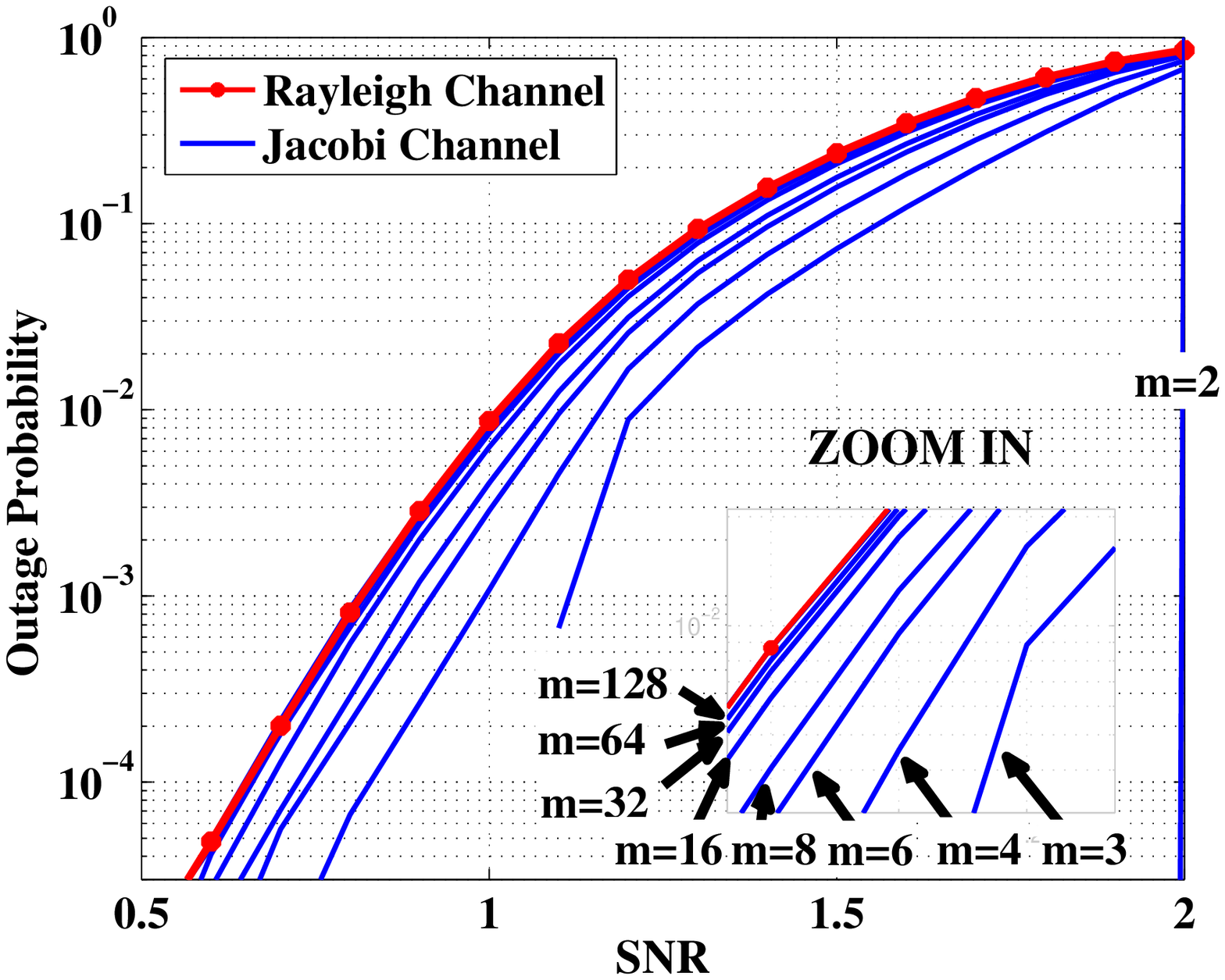,height=0.88\textwidth,width=1.1\textwidth,clip=}
\caption{} \label{Rayleigh_Jacobi_Figure_2}
\end{subfigure}
\caption{Comparing the $2\times 2$ Rayleigh and Jacobi models for various numbers of supported modes $m$. $\bar{\rho}$ is the average SNR at each receive antenna. The ergodic capacity is given in (\subref{Rayleigh_Jacobi_Figure_1}) and the outage probability for $\bar{\rho}=20$dB in (\subref{Rayleigh_Jacobi_Figure_2}).} \label{Rayleigh_Jacobi_Figure}
\end{figure}

The issue of the normalizing constant, $1/m$, should be further explained. With fixed $m_t,m_r$, increasing $m$ has two effects. One effect is power loss into the unaddressed modes. This effect is actually pretty strong, so that for a fixed $\rho$ the channel matrix, the received SNR, and hence the capacity vanish with $m$. The other effect, is that with increasing $m$ the channel matrix becomes more ``random'', e.g., the matrix elements becomes statistically independent, and so the model is closer to the Rayleigh model. To compare the Jacobi model to the Rayleigh mode, we need to compensate for the power loss with increasing $m$, and concentrate only on the ``randomness'' effect.
For this, we evaluate the channel characteristics (capacity, outage probability) in terms of $\bar{\rho}$, the average SNR at each receive mode, given by
\begin{equation}
\bar{\rho}=  \frac{\rho}{m_r}\mathbb{E}\|\mathbf{H}_{11}\|_F^2= \frac{\rho}{m_r}\mathbb{E}\sum_{i=1}^{m_{\text{min}}}\lambda_i~.
\end{equation}
In the Rayleigh channel $\mathbf{H}_{11}$ is Gaussian, thus ${\bar{\rho}}=\rho  m_t$. For the Jacobi channel $\bar{\rho}$ can be evaluated by applying the marginal PDF $f_{\lambda_i}(\lambda_i)$ of the channel's singular values. This PDF is computed in Appendix \ref{appendix_ergodicCapcaty}. Nonetheless, for $m \gg m_t,m_r$ we can apply Equations \eqref{approx_ray_1} and \eqref{approx_ray_2} to have $\bar{\rho}\approx \rho m_t/m$.

Following that,
in Fig. \ref{Rayleigh_Jacobi_Figure} we compare the Rayleigh and Jacobi models, for $m_t=m_r=2$. As $m$ increases, the Jacobi model approaches the Rayleigh model in terms of the ergodic capacity and outage probability, as a function of ${\bar{\rho}}$. For example, with this normalization, the difference between the ergodic capacities of the Rayleigh and Jacobi models is less than $0.1$dB already for $m\geq 32$.

\section{Discussion} \label{sec_discussion}
The Jacobi MIMO channel is defined by the transfer matrix $\mathbf{H}_{11}$, a truncated $m_r\times m_t$ portion of an $m\times m$  Haar distributed unitary matrix. By establishing the relation between the channel's singular values and the Jacobi ensemble of random matrices we derived the ergodic capacity, outage probability and optimal diversity-multiplexing tradeoff. An interesting phenomenon is observed when the parameters of the model satisfy $m_t+m_r>m$: for any realization of $\mathbf{H}_{11}$, $m_t+m_r-m$ singular values are $1$. This results in an ergodic capacity which is at least $m_t+m_r-m$ times the SISO capacity. In the non-ergodic scenario this results a promise for strictly zero outage probability and an exponentially decaying error probability (``infinite diversity'') for any transmission rate below $(m_t+m_r-m)\log(1+\rho)$.

The main motivation to define such a model comes from optical communication. Nonetheless, the results presented in this paper provide conceptual insights on fading channels in other communication scenaria, such as wireless communication. The size of the unitary matrix, $m$, can be viewed as the number of orthogonal propagation paths in the medium, whereas $m_t$ and $m_r$ are the number of addressed paths at the transmitter and receiver, respectively. The Jacobi fading model can be regarded as providing statistical model for the power loss in a system where for fixed $m_t$ and $m_r$, the size of the unitary matrix $m$ defines a {\emph{``fading measure''}} of the channel. For example, when $m$ is equal to $m_r$, the transfer matrix $\mathbf{H}_{11}$ is simply composed of orthonormal columns: its elements (i.e., the path gains) are highly dependent and there is no power loss at the receiver. As $m$ becomes greater, the orthogonality of the columns and rows of $\mathbf{H}_{11}$ fades, the dependency between the path gains becomes weaker and the power loss in the unaddressed receive outputs increases. Indeed, when $m$ is very large with respect to $m_t$ and $m_r$, with proper normalization that compensates for the power loss, the Jacobi fading model approaches to the Rayleigh model.

To conclude, the Jacobi model introduces new concepts in fading channels, providing a degree of freedom to scale the model from a unitary channel up to the Rayleigh channel, and therefore it may be of relevance, for example, in certain scenaria of wireless communication, where the worst case assumption of Rayleigh fading does not fit well the real behavior of the channel.

\section{Acknowledgement}
We wish to thank Amir Dembo and Yair Yona for interesting discussions on Lemma \ref{lem1}.

\appendices

\section{Proof of Theorem \ref{ergodicCap_int}} \label{appendix_ergodicCapcaty}
According to \eqref{eq_ergodicCapacity}, the ergodic capacity satisfies
\begin{align}
C({\text{\small{$m_t,m_r,m;\rho$}}})&=\mathbb{E}[\log \det(\text{I}_{m_t}+\rho  \mathbf{H}_{11}^\dag\mathbf{H}_{11})] \\
&=\mathbb{E}[\sum_{i=1}^{m_t}\log (1+\rho \lambda_i)] \label{regodicCap_1}
\end{align}
where we denote by $\{\lambda_i\}_{i=1}^{m_t}$ the eigenvalues of $\mathbf{H}_{11}^\dag\mathbf{H}_{11}$. To simplify notations let us assume $m_t\leq m_r$ (one can simply replace $m_t$ with $m_r$ to obtain the proof for $m_t>m_r$). Thus, we can write
\begin{align}
C({\text{\small{$m_t,m_r,m;\rho$}}})&=m_t\mathbb{E}[\log (1+\rho \lambda_1)]. \label{regodicCap_2}
\end{align}
Now, the joint distribution of the ordered eigenvalues $f_{\lambda}(\lambda_1,\ldots,\lambda_{m_t})$ is given by \eqref{JacobiPDF}. The joint distribution of the {\emph{unordered}} eigenvalues equals \[\tfrac{1}{m_t!}f_{\lambda}(\lambda_1,\ldots,\lambda_{m_t})~,\]
thus we can compute the density of $\lambda_1$ by integrating out $\{\lambda_i\}_{i=2}^{m_t}$, that is
\begin{align}
f_{\lambda_1}(\lambda_1)&=\int_0^1\ldots \int_0^1 \tfrac{1}{m_t!}f_{\lambda}(\lambda_1,\ldots,\lambda_{m_t}) \prod_{i=2}^{m_t}d\lambda_i~.
\end{align}
By taking
\begin{equation}
\lambda_i=\tfrac{1}{2}(1-\tilde{\lambda}_i)
\end{equation}
we can write
\begin{align}
f_{\tilde{\lambda}_1}(\tilde{\lambda}_1)&=\int_{-1}^1\ldots \int_{-1}^1 f_{\tilde{\lambda}}(\tilde{\lambda}_1,\ldots,\tilde{\lambda}_{m_t}) \prod_{i=2}^{m_t}d\tilde{\lambda}_i~,
\end{align}
where
\begin{align}
f_{\tilde{\lambda}}(\tilde{\lambda}_1,\ldots,\tilde{\lambda}_{m_t})=\tilde{K}_{m_t,m_r,m}^{-1}\prod_{i=1}^{m_t}
(1-\tilde{\lambda}_i)^{\alpha}(1+\tilde{\lambda}_i)^{\beta}\prod_{i<j}(\tilde{\lambda}_i-\tilde{\lambda}_j)^2~, \label{PDF_tilde}
\end{align}
and $\alpha=m_r-m_t$, $\beta=m-m_r-m_t$. Now, the term
\[\prod_{1\leq i <j\leq m_t}(\tilde{\lambda}_i-\tilde{\lambda}_j)\]
is the determinant of the Vandermonde matrix
\begin{equation} \label{vandermonde}
\begin{bmatrix} 1 & \ldots & 1\\ \tilde{\lambda}_1 & \ldots & \tilde{\lambda}_{m_t}\\ \vdots & & \vdots\\ \tilde{\lambda}_1^{m_t-1} & \ldots & \tilde{\lambda}_{m_t}^{m_t-1}\end{bmatrix}~.
\end{equation}
With row operations we can transform \eqref{vandermonde} into the following matrix
\begin{equation} \label{vandermonde2}
\begin{bmatrix} P^{(\alpha,\beta)}_0(\tilde{\lambda}_1) & \ldots & P^{(\alpha,\beta)}_0(\tilde{\lambda}_{m_t})\\ \vdots & & \vdots\\ P^{(\alpha,\beta)}_{m_t-1}(\tilde{\lambda}_1) & \ldots & P^{(\alpha,\beta)}_{m_t-1}(\tilde{\lambda}_{m_t})\end{bmatrix}~.
\end{equation}
where $P^{(\alpha,\beta)}_n(x)$ are the Jacobi polynomials \cite[8.96]{tableOfIntegrals}. These polynomials form a complete orthogonal system in the interval $[-1,1]$ with respect to the weighting function $w(x)=(1-x)^\alpha (1+x)^\beta$, that is
\begin{equation} \label{orthogonalPol}
\int_{-1}^1 w(x) P^{(\alpha,\beta)}_n(x)P^{(\alpha,\beta)}_k(x) dx=a_{k,\alpha,\beta} \delta_{kn}~,
\end{equation}
where the coefficients $a_{k,\alpha,\beta}$ are given by
\begin{equation}
a_{k,\alpha,\beta}=\frac{2^{\alpha+\beta+1}}{2k+\alpha+\beta+1}{2k+\alpha+\beta \choose k} {2k+\alpha+\beta \choose k+\alpha}^{-1}~.
\end{equation}
Thus we can write
\begin{align}
\prod_{1\leq i <j\leq m_t}(\tilde{\lambda}_i-\tilde{\lambda}_j)=C_{m_t,m_r,m}\sum_{\sigma\in S_{m_t}} (-1)^{sgn(\sigma)}\prod_{i=1}^{m_t}P^{(\alpha,\beta)}_{\sigma(i)-1}(\tilde{\lambda}_i)~, \label{detVandermonde}
\end{align}
where $S_{m_t}$ is the set of all permutations of $\{1,\ldots,m_t\}$, $sgn(\sigma)$ denotes the signature of the permutation $\sigma$ and $C_{m_t,m_r,m}$ is a constant picked up from the row operations on the Vandermonde matrix \eqref{vandermonde}. By applying \eqref{detVandermonde} into \eqref{PDF_tilde} we get
\begin{align}
f_{\tilde{\lambda}}(\tilde{\lambda}_1,\ldots,\tilde{\lambda}_{m_t})={\text{\small{$\tilde{C}_{m_t,m_r,m}^{-1}$}}}\sum_{\sigma_1,\sigma_2\in S_{m_t}} (-1)^{sgn(\sigma_1)+sgn(\sigma_2)} \prod_{i=1}^{m_t}
(1-\tilde{\lambda}_i)^{\alpha}(1+\tilde{\lambda}_i)^{\beta}P^{(\alpha,\beta)}_{\sigma_1(i)-1}(\tilde{\lambda}_i)P^{(\alpha,\beta)}_{\sigma_2(i)-1}(\tilde{\lambda}_i)~.
\end{align}
Further integrating over $\{\tilde{\lambda}_i\}_{i=2}^{m_t}$ results
\begin{align}
f_{\tilde{\lambda}_1}(\tilde{\lambda}_1)&={\text{\small{$\tilde{C}_{m_t,m_r,m}^{-1}$}}}\sum_{\sigma_1,\sigma_2\in S_{m_t}} (-1)^{sgn(\sigma_1)+sgn(\sigma_2)} (1-\tilde{\lambda}_1)^{\alpha}(1+\tilde{\lambda}_1)^{\beta}\times \nonumber\\
&\qquad \qquad \qquad \qquad \times P^{(\alpha,\beta)}_{\sigma_1(1)-1}(\tilde{\lambda}_1) P^{(\alpha,\beta)}_{\sigma_2(1)-1}(\tilde{\lambda}_1)\prod_{i=2}^{m_t} a_{(\sigma_1(i)-1),\alpha,\beta} \delta_{\sigma_1(i)\sigma_2(i)}\\
&={\text{\small{$\tilde{C}_{m_t,m_r,m}^{-1}$}}}(m_t-1)!\sum_{k=0}^{m_t-1} (1-\tilde{\lambda}_1)^{\alpha}(1+\tilde{\lambda}_1)^{\beta}[P^{(\alpha,\beta)}_{k}(\tilde{\lambda}_1)]^2\prod_{i\neq k} a_{i,\alpha,\beta}\\
&=\frac{1}{m_t}\sum_{k=0}^{m_t-1} a_{k,\alpha,\beta}^{-1}[P^{(\alpha,\beta)}_{k}(\tilde{\lambda}_1)]^2(1-\tilde{\lambda}_1)^{\alpha}(1+\tilde{\lambda}_1)^{\beta}~,
\end{align}
where the first equality follows from \eqref{orthogonalPol} and thus implies that $\sigma_1(i)=\sigma_2(i)$ for all $i$. This results in the second equality while the third follows from \eqref{orthogonalPol} and the fact that $f_{\tilde{\lambda}_1}(\tilde{\lambda}_1)$ must integrate to unity. Turning back to $\lambda_1$ we get:
\begin{align}
f_{{\lambda}_1}({\lambda}_1)&=\frac{1}{m_t}\sum_{k=0}^{m_t-1} b_{k,\alpha,\beta}^{-1}\big({P^{(\alpha,\beta)}_{k}(1-2\lambda_1)}\big)^2\lambda_1^{\alpha}(1-\lambda_1)^{\beta}~,
\end{align}
where
\begin{equation}
b_{k,\alpha,\beta}=\frac{1}{2k+\alpha+\beta+1}{2k+\alpha+\beta \choose k} {2k+\alpha+\beta \choose k+\alpha}^{-1}~.
\end{equation}

\section{Proof of Theorem \ref{thm_DMT_firstCase}} \label{Proof of Theorem_thm_DMT_firstCase}
The outage probability for a transmission rate $R$ is
\begin{equation}
P_{out}\text{\small{$(m_t,m_r,m;R)$}}= \inf_{\substack{Q:~Q\succeq 0}}Pr\big[\log \det(\text{I}_{m_r}+\rho  \mathbf{H}_{11}Q\mathbf{H}_{11}^\dag)<R\big]~,
\end{equation}
where the minimization is over all covariance matrices $Q$ of the transmitted signal that satisfy the power constraints. As was already mentioned, since the statistics of $\mathbf{H}_{11}$ is invariant under unitary permutations, the optimal choice of $Q$, when applying constant {{per-mode}} power constraint, is simply the identity matrix. When imposing power constraint on the total power over all modes, we can take $Q=\text{I}_{m_t}$ if $\rho \gg 1$ since
\begin{equation}
P_{out}\text{\small{$(m_t,m_r,m;R)$}}\doteq Pr\big[\log \det(\text{I}_{m_r}+\rho  \mathbf{H}_{11}\mathbf{H}_{11}^\dag)<R\big]~, \label{expOutage}
\end{equation}
where we use $\doteq$ to denote \emph{exponential equality}, i.e., $f(\rho)\doteq \rho^d$ denotes
\begin{equation}
\lim_{\rho\rightarrow \infty}\frac{\log f(\rho)}{\log \rho}=d~.
\end{equation}
Eq. \eqref{expOutage} can be proved by picking $Q=\text{I}_{m_t}$ to derive an upper bound on the outage probability and $Q=m_t\text{I}_{m_t}$ to derive a lower bound. It can be easily shown that these bounds are exponentially tight (see \cite{DMT}), hence, in the scale of interest, we can take $Q=\text{I}_{m_t}$.

Now, let the transmission rate be $R=r\log(1+\rho)$ and without loss of generality, let us assume that $m_t\leq m_r$ (the outage probability is symmetric in $m_t$ and $m_r$). Since \[\log \det(\text{I}_{m_r}+\rho  \mathbf{H}_{11}\mathbf{H}_{11}^\dag)=\log \det(\text{I}_{m_t}+\rho  \mathbf{H}_{11}^\dag\mathbf{H}_{11})\] we can apply the joint distribution of the ordered eigenvalues of $ \mathbf{H}_{11}^\dag\mathbf{H}_{11}$ to write
\begin{align}
P_{out}\text{\small{$(m_t,m_r,m;r\log(1+\rho))$}}&\doteq K^{-1}_{m_t,m_r,m}\int_{\mathcal{B}}\prod_{i=1}^{m_t} \lambda_i ^{m_r-m_t}(1-\lambda_i)^{m-m_r-m_t}\prod_{i<j}(\lambda_i-\lambda_j)^2d \mathbf{\underline\lambda}~,
\end{align}
where $K_{m_t,m_r,m}$ is a normalizing factor and
\[\mathcal{B}=\big\{\mathbf{\underline\lambda} :~0\leq \lambda_1\leq \ldots \leq \lambda_{m_t} \leq 1,~ \prod_{i=1}^{m_t} (1+\rho \lambda_i)<(1+\rho)^r\big\}\]
is the set that describes the outage event. Letting
\begin{equation}
\lambda_i=\rho^{-\alpha_i}~
\end{equation}
for $i=1,\ldots,m_t$ allows us to write
{{\begin{align}
P_{out}\text{\small{$(m_t,m_r,m;r\log(1+\rho))$}}\doteq (\log\rho)^{m_t}&K^{-1}_{m_t,m_r,m}\int_{\mathcal{B}}\prod_{i=1}^{m_t} \rho^{-\alpha_i(m_r-m_t+1)} \\
& (1-\rho^{-\alpha_i})^{m-m_r-m_t}\prod_{i<j}(\rho^{-\alpha_i}-\rho^{-\alpha_j})^2d \mathbf{\underline\alpha}~.
\end{align}}}
Since \[1+\rho^{1-\alpha_i}\doteq \rho^{(1-\alpha_i)^+}~,\] where $(x)^+=\max\{0,x\}$, we can describe the set of outage events by
\[\mathcal{B}=\big\{\mathbf{\underline\alpha} :~\alpha_1\geq \ldots \geq \alpha_{m_t} \geq 0,~ \sum_{i=1}^{m_t} (1-\alpha_i)^+<r\big\}~.\]
Now, the term $(\log\rho)^{m_t}K^{-1}_{m_t,m_r,m}$ satisfies
\begin{equation}
\lim_{\rho\rightarrow \infty}\frac{\log ((\log\rho)^{m_t}K^{-1}_{m_t,m_r,m})}{\log \rho}=0~,
\end{equation}
thus we can write
\begin{align}
P_{out}\text{\small{$(m_t,m_r,m;r\log(1+\rho))$}}&\doteq \int_{\mathcal{B}}\prod_{i=1}^{m_t} \rho^{-\alpha_i(m_r-m_t+1)} \times\nonumber\\
& \qquad \qquad \qquad \times(1-\rho^{-\alpha_i})^{m-m_r-m_t}\prod_{i<j}(\rho^{-\alpha_i}-\rho^{-\alpha_j})^2d \mathbf{\underline\alpha}\\
&\leq \int_{\mathcal{B}}\prod_{i=1}^{m_t} \rho^{-\alpha_i(m_r-m_t+1)}\prod_{i<j}(\rho^{-\alpha_i}-\rho^{-\alpha_j})^2d \mathbf{\underline\alpha}~.
\end{align}
In \cite[Theorem 4]{DMT} it was shown that the right hand side of above satisfies
\begin{align}
\int_{\mathcal{B}}\prod_{i=1}^{m_t} \rho^{-\alpha_i(m_r-m_t+1)}\prod_{i<j}(\rho^{-\alpha_i}-\rho^{-\alpha_j})^2d \mathbf{\underline\alpha}\doteq \rho^{-f(\underline\alpha^*)}~,
\end{align}
where
\begin{equation}
f(\underline\alpha)=\sum_{i=1}^{m_t}(2i-1+m_r-m_t)\alpha_i
\end{equation}
and
\begin{equation}
\underline\alpha^*=\arg\inf_{\underline\alpha \in \mathcal{B}} f(\underline\alpha)~.
\end{equation}
By defining $S_\delta=\{\underline\alpha:~\alpha_i>\delta ~\forall~i=1,\ldots,m_t\}$ for any $\delta>0$, we can write
\begin{align}
P_{out}\text{\small{$(m_t,m_r,m;r\log(1+\rho))$}}&\geq \int_{\mathcal{B}\bigcap S_\delta}\prod_{i=1}^{m_t} \rho^{-\alpha_i(m_r-m_t+1)} \times \nonumber\\
&\qquad \qquad \times (1-\rho^{-\alpha_i})^{m-m_r-m_t}\prod_{i<j}(\rho^{-\alpha_i}-\rho^{-\alpha_j})^2d \mathbf{\underline\alpha}\\
&\geq (1-\rho^{-\delta})^{m_t(m-m_r-m_t)}\int_{\mathcal{B}\bigcap S_\delta}\prod_{i=1}^{m_t} \rho^{-\alpha_i(m_r-m_t+1)} \times\nonumber\\
&\qquad \qquad  \times\prod_{i<j}(\rho^{-\alpha_i}-\rho^{-\alpha_j})^2d \mathbf{\underline\alpha}\\
&\doteq \rho^{-f(\underline\alpha_\delta^*)}~,
\end{align}
where
\begin{equation}
\underline\alpha_\delta^*=\arg\inf_{\underline\alpha \in \mathcal{B}\bigcap S_\delta} f(\underline\alpha)~.
\end{equation}
Using the continuity of $f$, $\underline\alpha_\delta^*$ approaches $\underline\alpha^*$ as $\delta$ goes to zero and we can conclude that
\begin{align}
P_{out}\text{\small{$(m_t,m_r,m;r\log(1+\rho))$}}\doteq \rho^{-f(\underline\alpha^*)}~. \label{finalOutProof}
\end{align}
This result was obtained in \cite{DMT} for the Rayleigh model. From here one can continue as was presented in \cite{DMT}, showing that the error probability is dominated by the outage probability at high SNR (large $\rho$) for $l\geq m_t+m_r-1$ (\cite[Lemma 5 and Theorem 2]{DMT}, these proofs rely on \eqref{finalOutProof} without making any assumptions on the channel statistics, therefore are true also for the Jacobi model).

% reference http://journals.cambridge.org/action/displayFulltext?type=1&fid=298727&jid=ANU&volumeId=14&issueId=-1&aid=298726
% reference DISTRIBUTIONS OF THE EXTREME EIGENVALUES OF BETA–JACOBI RANDOM MATRICES: http://www-math.mit.edu/~plamen/files/kd.PDF
% reference: The beta-Jacobi matrix model, the CS decomposition, and generalized singular value problems, Alan Edelman ,  Brian ,  D. Sutton
% reference: http://web.mit.edu/18.338/www/Acta05rmt.PDF
% reference: Capacity of Multi-antenna Gaussian Channels
% reference: A. M. Tulino and S. Verd´u, Random matrix theory and wireless communications (now Publishers Inc., Hanover, 2004).
% reference: R. J. Muirhead, Aspects of multivariate statistical theory (Wiley, 1982).
% reference: Moments of Wishart-Laguerre and Jacobi ensembles of random matrices: application to the quantum transport problem in chaotic cavities
%reference: A matrix model for the ß-Jacobi ensemble

\bibliographystyle{IEEEtran}
\bibliography{IEEEabrv,mybib2}

\begin{thebibliography}{10}
\providecommand{\url}[1]{#1}
\csname url@samestyle\endcsname
\providecommand{\newblock}{\relax}
\providecommand{\bibinfo}[2]{#2}
\providecommand{\BIBentrySTDinterwordspacing}{\spaceskip=0pt\relax}
\providecommand{\BIBentryALTinterwordstretchfactor}{4}
\providecommand{\BIBentryALTinterwordspacing}{\spaceskip=\fontdimen2\font plus
\BIBentryALTinterwordstretchfactor\fontdimen3\font minus
  \fontdimen4\font\relax}
\providecommand{\BIBforeignlanguage}[2]{{%
\expandafter\ifx\csname l@#1\endcsname\relax
\typeout{** WARNING: IEEEtran.bst: No hyphenation pattern has been}%
\typeout{** loaded for the language `#1'. Using the pattern for}%
\typeout{** the default language instead.}%
\else
\language=\csname l@#1\endcsname
\fi
#2}}
\providecommand{\BIBdecl}{\relax}
\BIBdecl

\bibitem{Muirhead}
R.~J. Muirhead, \emph{Aspects of Multivariate Statistical Theory}.\hskip 1em
  plus 0.5em minus 0.4em\relax New York: Wiley, 1982.

\bibitem{MehtaRandomMatrices}
M.~L. Mehta, \emph{Random Matrices}.\hskip 1em plus 0.5em minus 0.4em\relax 3rd
  ed. New York: Academic Press, 1991.

\bibitem{EdelmanRandomMatrixTheory}
A.~Edelman and N.~R. Rao, ``Random matrix theory,'' \emph{Acta Numerica},
  vol.~14, pp. 233--297, 2005.

\bibitem{foschiniWirelessMIMO}
G.~J. Foschini, ``Layered space-time architecture for wireless communication in
  a fading environment when using multi-element antennas,'' \emph{Bell Labs
  Technical Journal}, vol.~1, no.~2, pp. 41--59, 1996.

\bibitem{TelatarCapacity}
I.~E. Telatar, ``Capacity of multi-antenna gaussian channels,'' \emph{European
  Transactions on Telecommunications}, vol.~10, pp. 585--595, 1999.

\bibitem{TulinoRandomMatrixTheory}
A.~M. Tulino and S.~Verd\'{u}, ``Random matrix theory and wireless
  communications,'' \emph{Commun. Inf. Theory}, vol.~1, pp. 1--182, June 2004.

\bibitem{OntheCapacityOfMultipleAntennaSystemsInRicianFading}
S.~Jayaweera and H.~Poor, ``On the capacity of multiple-antenna systems in
  rician fading,'' \emph{IEEE Transactions on Wireless Communications}, vol.~4,
  no.~3, pp. 1102 -- 1111, may 2005.

\bibitem{CapacityOfMIMORicianChannels}
M.~Kang and M.~Alouini, ``Capacity of mimo rician channels,'' \emph{IEEE
  Transactions on Wireless Communications}, vol.~5, no.~1, pp. 112 -- 122, jan.
  2006.

\bibitem{DigitalCommunicationsOverFadingChannels}
M.~K. Simon and M.~S. Alouini, \emph{Digital Communications Over Fading
  Channels}.\hskip 1em plus 0.5em minus 0.4em\relax New York: Wiley, 2000.

\bibitem{Nakagami}
M.~Nakagami, ``The $m$-distribution - a general formula of intensity
  distribution of rapid fading,'' \emph{Statistical Methods in Radio Wave
  Propagation. New York: Pergamon, 1960}, pp. 3--–36.

\bibitem{capacityOfNakagamiChannel}
G.~Fraidenraich, O.~Leveque, and J.~M. Cioffi, ``On the mimo channel capacity
  for the nakagami-$m$ channel,'' \emph{IEEE Transactions on Information
  Theory}, vol.~54, no.~8, pp. 3752 --3757, aug. 2008.

\bibitem{RandomMatrixModelforNakagamiHoytFading}
S.~Kumar and A.~Pandey, ``Random matrix model for nakagami-hoyt fading,''
  \emph{IEEE Transactions on Information Theory}, vol.~56, no.~5, pp. 2360
  --2372, may 2010.

\bibitem{ScalingOpticalCommunications}
R.~W. Tkach, ``Scaling optical communications for the next decade and beyond,''
  \emph{Bell Labs Technical Journal}, vol.~14, no.~4, pp. 3--10, 2010.

\bibitem{capacityCrunch}
A.~R. Chraplyvy, ``The coming capacity crunch,'' \emph{European Conference on
  Optical Communication (ECOC)}, plenary talk, 2009.

\bibitem{CapacityScalingThroughSpatialMultiplexing}
P.~Winzer, ``Energy-efficient optical transport capacity scaling through
  spatial multiplexing,'' \emph{Photonics Technology Letters, IEEE}, vol.~23,
  no.~13, pp. 851 --853, july1, 2011.

\bibitem{NewGenerationOpticalInfrastructureTechnologies}
T.~Morioka, ``New generation optical infrastructure technologies: Exat
  initiative towards 2020 and beyond,'' in \emph{OptoElectronics and
  Communications Conference (OECC)}, 2009.

\bibitem{WinzerFoschiniOpticalMIMO}
P.~J. Winzer and G.~J. Foschini, ``Mimo capacities and outage probabilities in
  spatially multiplexed optical transport systems.'' \emph{Optics Express},
  vol.~19, no.~17, pp. 16\,680--96, 2011.

\bibitem{TheUnderAddressedChannel}
R.~Dar, M.~Feder, and M.~Shtaif, ``The underaddressed optical multiple-input,
  multiple-output channel: capacity and outage,'' \emph{Optics Letters},
  vol.~37, no.~15, pp. 3150--3152, 2012.

\bibitem{EdelmanCSdecomposition}
A.~Edelman and B.~D. Sutton, ``The beta-jacobi matrix model, the cs
  decomposition, and generalized singular value problems,'' \emph{Foundations
  of Computational Mathematics}, vol.~8, no.~1, pp. 259--285, 2008.

\bibitem{tableOfIntegrals}
I.~S. Gradshteyn and I.~M. Ryzhik, \emph{Table of Integrals, Series, and
  Products}.\hskip 1em plus 0.5em minus 0.4em\relax New York: Academic Press,
  1980, vol.~48.

\bibitem{DMT}
L.~Zheng and D.~N.~C. Tse, ``Diversity and multiplexing: A fundamental tradeoff
  in multiple antenna channels,'' \emph{IEEE Trans. Inform. Theory}, vol.~49,
  pp. 1073--1096, 2002.

\bibitem{Alamouti}
S.~Alamouti, ``A simple transmit diversity technique for wireless
  communications,'' \emph{Selected Areas in Communications, IEEE Journal on},
  vol.~16, no.~8, pp. 1451 --1458, oct 1998.

\end{thebibliography}

\end{document}